\documentclass[%
aps,
reprint,
superscriptaddress,
 amsmath,amssymb, aps,
prl,
floatfix
]{revtex4-1}

\usepackage{color,soul}
\usepackage{graphicx}
\usepackage{dcolumn}
\usepackage{bm}
\usepackage{xcolor}
\usepackage{xr}

\begin{document}
\setlength{\abovedisplayskip}{0pt}
\setlength{\belowdisplayskip}{0pt}
\setlength{\headsep}{0pt}
\setlength{\partopsep}{0pt}


\title{Evolution of Helimagnetic Correlations when approaching the \\Quantum Critical Point of Mn$_{1-x}$Fe$_x$Si}

\author{C. Pappas}
\affiliation{Faculty of Applied Sciences, Delft University of Technology, Mekelweg 15, 2629 JB Delft, The Netherlands}
\author{A. O. Leonov}
\affiliation{Chiral Research Center, Hiroshima University, Higashi Hiroshima, Hiroshima 739-8526, Japan}
 \email{c.pappas@tudelft.nl}
\author{L.J. Bannenberg}
\affiliation{Faculty of Applied Sciences, Delft University of Technology, Mekelweg 15, 2629 JB Delft, The Netherlands}
\author{P. Fouquet}
\affiliation{Institut Laue Langevin, 71 Avenue des Martyrs, CS 20156, Grenoble, France}
\author{T. Wolf}
\affiliation{Institute for Solid State Physics, Karlsruhe Institute of Technology, 76131 Karlsruhe, Germany}
\author{F. Weber}
\affiliation{Institute for Solid State Physics, Karlsruhe Institute of Technology, 76131 Karlsruhe, Germany}
\affiliation{Institute for Quantum Materials and Technologies, Karlsruhe Institute of Technology, 76131 Karlsruhe, Germany}
\date{\today}

\begin{abstract}

We present a comprehensive investigation of the evolution of helimagnetic correlations in Mn$_{1-x}$Fe$_x$Si  with increasing doping. By combining  polarised neutron scattering and high resolution Neutron Spin Echo spectroscopy we investigate three samples with $x$=0.09, 0.11  and 0.14, i.e. with compositions on both sides of  the concentration $x^*\sim 0.11$ where the helimagnetic Bragg peaks disappear and between $x^*$ and the quantum critical concentration $x_C \sim 0.17$, where $T_C$ vanishes. We find that the abrupt disappearance of the long range helical periodicity at $x^*$, does not affect the precursor fluctuating correlations. These build up  with decreasing temperature in a similar way as for the parent compound MnSi. Also the dynamics bears strong similarities to MnSi. The analysis of our results indicates that  frustration, possibly due to  achiral RKKY interactions,  increases with increasing  Fe doping. We argue that this effect  explains both the expansion of the precursor phase with increasing $x$ and the abrupt disappearance of long range helimagnetic periodicity at $x^*$.

\end{abstract}

\maketitle

\section{Introduction}

The physics of the chiral magnet MnSi touches onto several fundamental questions in condensed-matter physics, from the stabilisation of exotic  states like chiral skyrmions \cite{muhlbauer2009} to the interplay between localised and itinerant magnetism \cite{ishikawa1977, Thessieu1995, Corti:2007kx, demishev2011, Yasuoka:2016hg, Bannenberg2019pressure, Yaouanc2020} as well as to non-Fermi liquid behaviour \cite{pfleiderer2001, doiron2003, pfleiderer2007,  lee2009, ritz2013, Kirkpatrick2018}  and quantum fluctuations \cite{pfleiderer1997, BlueQuantumFog2006, rossler2006, Hopkinson2009, Pfleiderer2009dr, KirkpatrickPRL2010, KirkpatrickPRB2010, Kruger2012, Povzner2018}.   Under pressure a non-Fermi liquid behaviour sets-in without quantum criticality  \cite{pfleiderer1997, pfleiderer2004,pintschovius2004, pfleiderer2007}. Furthermore,  the first order transition temperature $T_C$ is driven to 0~K at $p_C\sim 1.4$~GPa, although the  magnetic moment does not vanish. In the region of the temperature-pressure phase diagram, where $p \gtrsim p_C$ and $T_C=0$ K, long range range spiral and skyrmion correlations are restored under  magnetic fields  \cite{Bannenberg2019pressure}, a result that has been attributed to a softening of the  magnetic moment.  Pressure would therefore  enhance the itinerant electron character of magnetism triggering the suppression of $T_C$. On the other hand, in the absence of  quantum critical point (QCP) at $p_C$ \cite{Pfleiderer2009dr},  the origin and nature of the highly debated non-Fermi liquid phase  \cite{rossler2006, BlueQuantumFog2006, Binz2006,  Fischer2008, Hopkinson2009, KirkpatrickPRL2010} remains an open question. It was suggested that this phase, out of which magnetic fields induce long range spiral correlations is  fluctuating and possibly of quantum nature \cite{Pfleiderer2009dr}. However, our high resolution Neutron Spin Echo (NSE) spectroscopy measurements did not reveal the existence of such fluctuations, possibly  due to  limitations (background contribution of the pressure cell) inherent to measurements under high pressures  \cite{Bannenberg2019pressure}.

\begin{figure}[tb]
\begin{center}
\includegraphics[width= 0.38\textwidth]{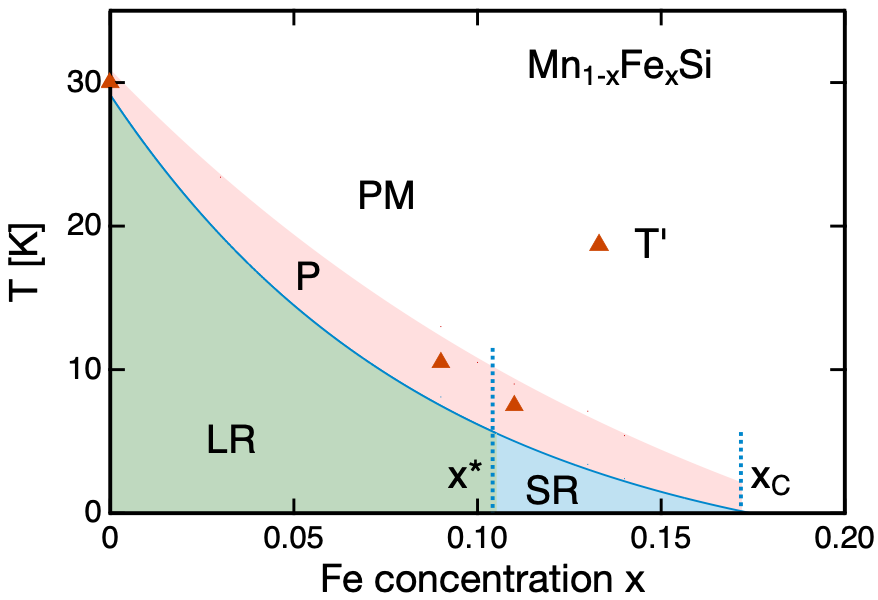}
\caption{Phase diagram of Mn$_{1-x}$Fe$_x$Si deduced from our previous SANS and susceptibility measurements  \cite{bannenberg2018mnfesisans, bannenberg2018mnfesisquid}.  The pink shaded area indicates the precursor phase, P, 
and PM stands for the paramagnetic phase. The blue line indicates the transition to the ordered helimagnetic phase with long (LR) and short (SR) range periodicity respectively. The temperatures $T^\prime$ have been determined as shown in Fig. \ref{intensity_chirality} (b) and (c).
}
\label{Phase_diagram}
\end{center}
\end{figure}

These limitations are  overcome by  chemical pressure, in the form of Fe doping in Mn$_{1-x}$Fe$_x$Si.  The behaviour of this system resembles that of  MnSi under pressure  \cite{bauer2010, grigoriev2009b, grigoriev2011, franz2014, bannenberg2018mnfesisans, bannenberg2018mnfesisquid, Petrova_2019_MnFeSi,  demishev2014, demishev2016a, KindervaterPRB2020} and our magnetization, susceptibility and SANS investigations \cite{bannenberg2018mnfesisans, bannenberg2018mnfesisquid}  led to the phase diagram shown in Fig. \ref{Phase_diagram}.  With increasing doping, the transition temperature decreases continuously and vanishes at $x_C\sim$0.17.  On the other hand, the helimagnetic Bragg peaks, a signature of long range helimagnetic periodicity, disappear abruptly at a much lower concentration of $x^*\sim$ 0.11. For $x\gtrsim x^*$ magnetic susceptibility and electric transport phenomena reveal a non-Fermi liquid behaviour, which, as in MnSi under pressure, has been attributed to a chiral spin liquid state governed by quantum fluctuations  \cite{demishev2014, demishev2016a, KindervaterPRB2020}.  Here we investigate  the evolution of helimagnetic correlations and their dynamics as a function of chemical substitution in Mn$_{1-x}$Fe$_x$Si 
using polarised neutron scattering and Neutron Spin Echo (NSE)  spectroscopy.  We investigated three samples with $x$=0.09, 0.11  and 0.14, i.e. with compositions on both sides of  $x^*$ and between $x^*$ and $x_C$. Our results reveal  that the abrupt disappearance of  long range helimagnetic periodicity at $x^*$ does not affect the  precursor fluctuating correlations. These build up with decreasing temperature in a way that is very similar to that of the  parent compound MnSi. Also the dynamics, i.e. the characteristic relaxation times and their temperature dependence, resembles the behaviour of MnSi.

In order to understand these results we discuss the evolution of magnetic interactions with doping and also compare Mn$_{1-x}$Fe$_x$Si with MnSi under pressure. Our analysis brings us to the conclusion that in Mn$_{1-x}$Fe$_x$Si  doping introduces frustration. We argue that this effect  explains both the  destabilisation of  long range helimagnetic periodicity at $x^*$ as well as the robustness of the precursor phase and its expansion with increasing $x$. 

\begin{figure}[tb]
\begin{center}
\includegraphics[width= 0.38\textwidth]{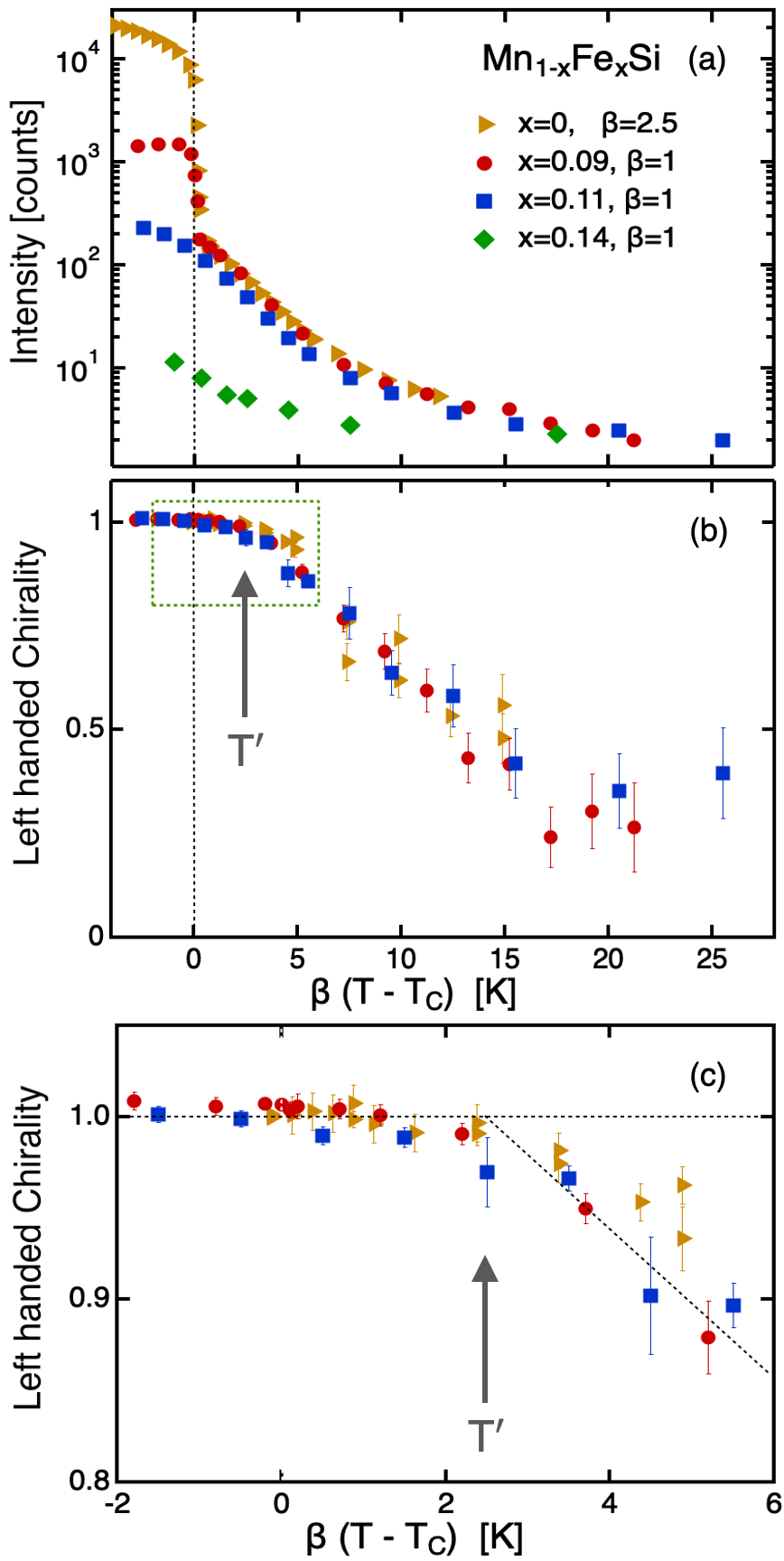}
\caption{Temperature dependence of the helimagnetic SANS intensity (a) and the left handed chiral fraction of the scattering (b) and (c). For the sake of comparison between the results of the three compositions investigated here and the parent compound MnSi,  the abscissa is the scaled temperature difference $\beta\; (T - T_C )$, with $\beta$=2.5 for MnSi and $\beta$ = 1 otherwise. In this way is is possible to account for the broadening of the precursor phase found for the doped samples. $T_C$=7.8, 5 and 2.5 K for  $x$=0.09, 0.11 and 0.14 respectively. Panel (c) shows a close-up view of the green rectangle area shown in (b). The  dotted lines in (c) illustrate the determination of at  $T^\prime \sim \beta\; (T - T_C )$=2.5. For $T <  T^\prime$ the  scattering of MnSi, Mn$_{0.91}$Fe$_{0.09}$Si and Mn$_{0.89}$Fe$_{0.11}$Si reaches full left handed chirality within at least two times the error bars.   }
\label{intensity_chirality}
\end{center}
\end{figure}
\begin{figure}[tb]
\begin{center}
\includegraphics[width= 0.4\textwidth]{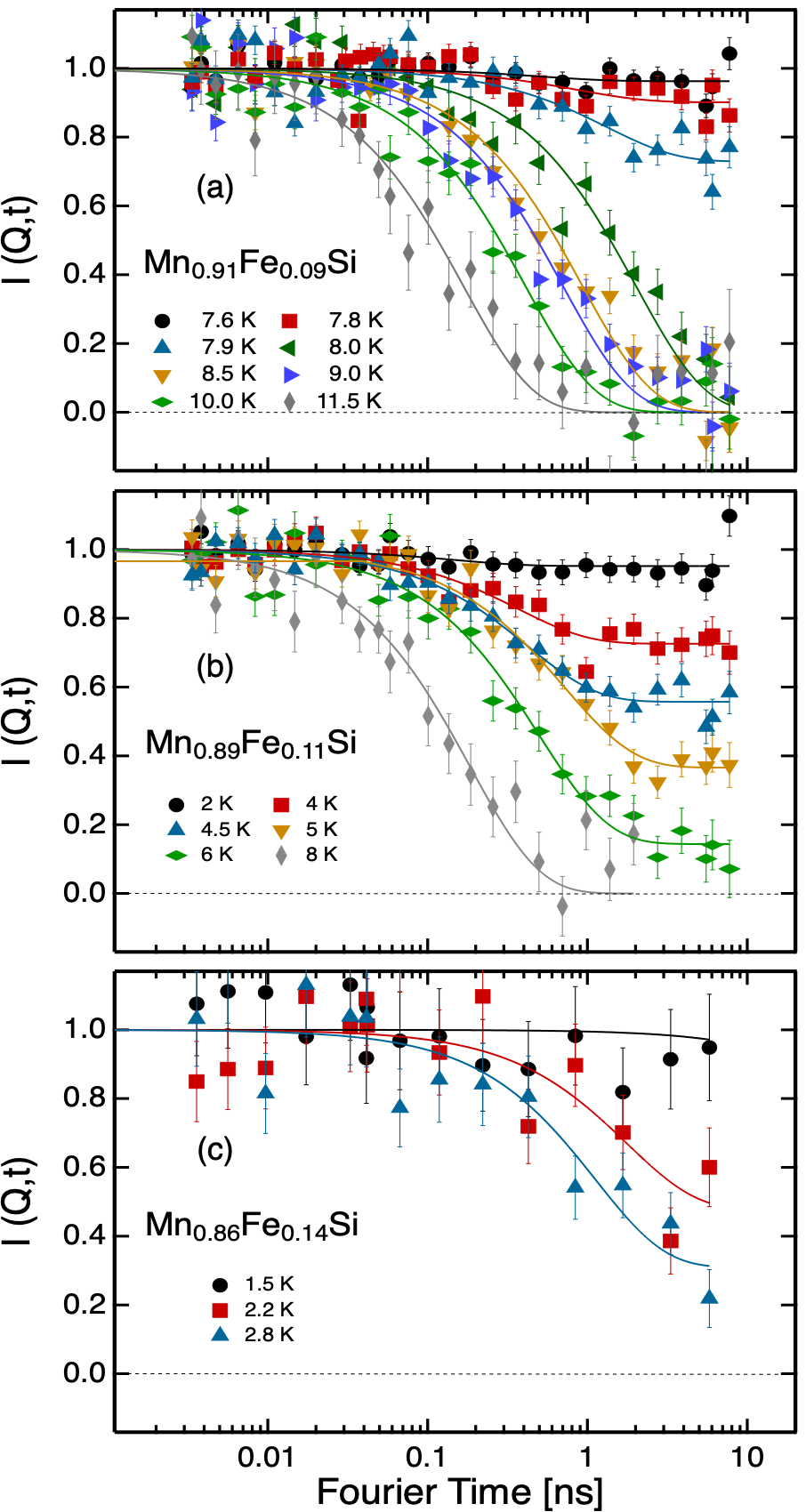}
\caption{ Intermediate scattering functions, determined by Neutron Spin Echo spectroscopy,  of Mn$_{0.91}$Fe$_{0.09}$Si (a), Mn$_{0.89}$Fe$_{0.11}$Si (b) and Mn$_{0.86}$Fe$_{0.14}$Si (c). The lines represent fits to an exponential decay (see text).   }
\label{NSE_spectra}
\end{center}
\end{figure}

\begin{figure*}[htb]
\begin{center}
\includegraphics[width= \textwidth]{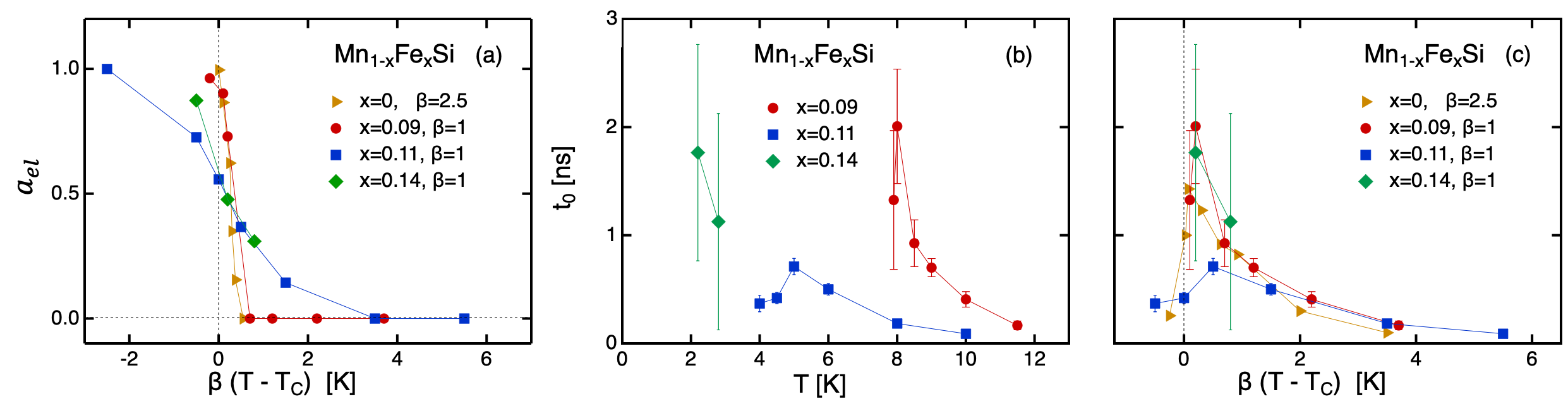}
\caption{Evolution with temperature and doping of the  elastic fractions $a_{el}$ and the characteristic times $t_0$  deduced from fitting an exponential decay to the NSE spectra of Fig. \ref{NSE_spectra} (see text). In (a) and (c)  the evolution of $a_{el}$ and $t_0$  is compated with that of parent compound MnSi and for this reason the abscissa is, as in Fig. \ref{intensity_chirality}, the scaled temperature difference $\beta\; (T - T_C )$, with $\beta$=2.5 for MnSi and $\beta$ = 1 otherwise.  }
\label{tau_el}
\end{center}
\end{figure*}

\section{Experimental Results}

The  samples were grown using the Bridgeman method. For  $x$=0.09, 0.11 we measured the same single crystals as in our previous SANS study \cite{bannenberg2018mnfesisans}. For x=0.14 and in order to compensate for the neutron intensity losses at this high Fe doping we chose a large polycrystalline and racemate polycrystalline sample, from the same batch as the samples of our previous studies \cite{bannenberg2018mnfesisans, bannenberg2018mnfesisquid}. On one hand, this choice is not problematic, since at this composition the  magnetic Bragg peaks have disappeared and the magnetic correlations are no longer pinned to the lattice. On the other hand, as it will be explained below, the polarised neutron scattering from this sample cannot be used to determine the degree of chirality of the magnetic correlations. 

The measurements were performed at the IN11 spectrometer, of the Institut Laue Langevin, using the paramagnetic NSE configuration and  an incoming wavelength of 0.55 nm. For each sample we determined the polarised neutron scattering and the NSE spectra at the respective maxima of the magnetic scattering intensity. These occur at the scattering vector values $Q$=0.58, 0.68 and 0.88 nm$^{-1}$ leading to helical pitches  $\ell$=10.8, 9.2 and 7.1 nm, for x=0.09, 0.11 and 0.14 respectively,  in good agreement with our previous SANS results  \cite{bannenberg2018mnfesisans}.

By exploiting the polarisation analysis capabilities of the experimental set-up  we obtained an accurate determination of the magnetic scattering and of the chiral fraction of the magnetic correlations  \cite{Blume, Maleyev2, Murani:1980vg, PAPPAS2006521, pappas2009, pappas2011}. The results, shown in 
 Fig. \ref{intensity_chirality}(b) and (c), highlight the  abrupt change of behaviour at  $x^*$. 


For $x=0.09$, i.e. for $x<x^*$, 
 a jump of  intensity 
marks the onset of the helimagnetic Bragg peaks and the first order phase transition at $T_C=7.8 
$ K. 
This  jump disappears for $x>x^*$. However, for $x$=0.11 the  precursor correlations build up  with decreasing temperature in the same way as for $x$=0.09. Consequently,  the precursor phase builds up in  a similar way on both sides of $x^*$, 
 a result that is hard to reconcile with the Brazovskii scenario proposed to explain the  first order phase transition in MnSi and other chiral magnets  \cite{janoschek2013}.  
 For $x$=0.14, the evolution of intensity with temperature is much slower and can only be superimposed with data from lower dopings assuming a negative $T_C$. 
 
Fig. \ref{intensity_chirality} (b) and (c)  shows that the scattering from both single crystalline samples, with $x=0.09$ and 0.11, is fully chiral at low temperatures. 
Unpublished results on single crystals indicate that the correlations remain chiral for dopings  even higher than $x_C$ \cite{Grigoriev_private}.  However, the scattering from our 14\% polycrystalline sample was achiral. This brought us to the conclusion  that it is a racemate combining grains with different structural chirality, because in this system structural and magnetic chiralities are coupled \cite{Grigoriev2010_MnFeSi_chiral}. Thus, the polarised neutron scattering from this sample does not reflect the chirality of  magnetic correlations and for this reason this concentration is not included in Fig. \ref{intensity_chirality} (b) and (c).

For $x=0.09$ and 0.11, full chirality extends  up to  $T^\prime \sim T_C+2.5$~K, thus well above $T_C$.  
The temperature interval $T^\prime -T_C $ is  $\sim2.5$ times broader than in MnSi, where full chirality is found up to $T^\prime \sim  T_C+1$~K   \cite{pappas2009, pappas2011}.  
In order to account for this difference when comparing the behaviour of the pristine and doped samples, the abscissas in Fig. \ref{intensity_chirality} are the scaled temperature differences $\beta \, (T-T_C)$, with $\beta$=2.5 for MnSi and $\beta=1$ otherwise. In these plots Mn$_{0.91}$Fe$_{0.09}$Si reproduces the behaviour of MnSi, for both the intensity and the chiral fraction.

Further insight in the effect of chemical doping on the magnetic behaviour is provided by the intermediate scattering functions $I(Q=\tau, t)$, determined by NSE spectroscopy and shown in Fig. \ref{NSE_spectra}. The spectra follow an exponential decay that can be fitted by the function $I(Q=\tau, t) = (1-a_{el}) \exp(-t/t_0) + a_{el}$, where $t_0$ is the characteristic relaxation time and $a_{el}$ the elastic fraction of the scattering. This behaviour  contrasts with that of other disordered helimagnets, such as Fe$_{0.7}$Co$_{0.3}$Si \cite{bannenberg2017, bannenberg2016squid} or Zn doped Cu$_2$OSeO$_3$ \cite{Birch2019dw}, where strong deviations from exponentiality have been reported. 

The exponential relaxation rules out a  spin-glass-like ground state, the footprint of which would have been a stretched exponential  decay \cite{Pickup2009, Pappas2003}. 
Furthermore, the NSE spectra become completely elastic at the base temperature, 
a result, which excludes the spin liquid  scenario  suggested for $x>x^*$ \cite{demishev2014, demishev2016a, KindervaterPRB2020}. The elastic fraction, depicted in  Fig. \ref{tau_el} (a), reflects the evolution of the scattered intensity shown in Fig.\ref{intensity_chirality}(b). The change at $T_C$  is  almost step-like,   characteristic of the first order phase transition,  for MnSi and Mn$_{0.91}$Fe$_{0.09}$Si. On the other hand,  for  $x>x^*$, $a_{el}$  increases  gradually with decreasing temperature as also found in the disordered helimagnet Fe$_{0.7}$Co$_{0.3}$Si  \cite{bannenberg2017}. 

The  characterisitic relaxation times, $t_0$, depicted in Fig. \ref{tau_el} (b)-(c), 
do not change with doping and are comparable to those of MnSi \cite{pappas2009, pappas2011}.  Their values vary  between  0.1 and 2~ns, leading to  characteristic energies,  $\hbar \omega$, between 6.58~$\mu$eV and 0.33 $\mu$eV. These values are much lower than those  reported by a previous study \cite{grigoriev2011}, which however suffered from a low Q and energy resolution.   
The energies found here correspond to  temperatures between 80 and 4~mK. Consequently, 
these are  classical fluctuations, with $\hbar \omega \ll kT$,  not  the  quantum fluctuations  discussed in the literature \cite{demishev2014, demishev2016a, KindervaterPRB2020}. Thus,  classical critical slowing down prevails the dynamics of  helimagnetic correlations for $x>x^*$  masking the quantum criticality associated with the putative QCP at $x_C$.  


\section{Discussion}

Our results show that Fe doping affects the first order transition and the long range helical periodicity in a very different way than  the precursor phase. While the former disappear abruptly at $x^*$,  the latter expands and persists up to  highier dopings. Also the characteristic relaxation times of the fluctuations are comparable to those found in MnSi. In order to understand these results and in particular the robustness of the precursor phase,  we adopt the Dzyaloshinskii model for cubic non-centrosymmetric ferromagnets \cite{Dzyaloshinskii1964a, Dzyaloshinskii1965a, bak1980}, which leads to the free energy per unit cell \cite{qian2018}:
\begin{equation}
{\cal E} = 
\frac{Ja^2}{2} \sum_{i=x,y,z}\partial_i \bm m \cdot \partial_i \bm m +
Da\;  \bm m \cdot \bm \nabla \times \bm m - a^3  \mu_0 M \bm m \cdot {\bm H} + {\cal E}_{a},
\label{eq:model}
\end{equation} 
with $J$ the ferromagnetic Heisenberg exchange, $D$ the Dzyaloshinskii-Moriya (DMI) interaction,  $\bm m$  the unit vector in the direction of the magnetization, $M$  the magnetization, 
$a$ the lattice constant, and ${\cal E}_{a}$  the magnetic anisotropy energy. The ground state consists of helical spirals with a periodicity of $\ell =2 \pi a \, J / D$ propagating along specific crystallographic directions fixed by  magnetic anisotropy.  A magnetic field strong enough to overcome the anisotropy, aligns the spirals   towards its direction and  tilts the magnetic moments by an angle given by $\cos \theta = {H}/{H_{C2}}$, where  $H_{C2}$ is the critical field above which the homogeneous or field-polarised state sets-in.  The model of Eq. \ref{eq:model} leads to $\mu_0 M_0 H_{C2} = D^2/J a^3$, with $M_0$ the volume magnetization  at $H_{C2}$.  Therefore, at low temperatures,  the pitch of the helical spirals can be written as: $\ell = 2 \pi a \sqrt{ J/ (a^3 \, \mu_0 M_0 H_{C2} )}$ and by assuming that  $J$ is given by  the Curie-Weiss temperature, $T_{CW}$ (mean field approximation), 
it can be derived from   quantities that have been  determined experimentally (see e.g. Table 1 of \cite{bannenberg2018mnfesisquid}). In particular the  lattice constant, $a$,  does not change significantly with temperature \cite{pfleiderer2007} but varies  linearly with $x$, between 0.4565~nm for MnSi and 0.449~nm for FeSi  \cite{Grigoriev2010_MnFeSi_chiral}.

\begin{figure}[tb]
\begin{center}
\includegraphics[width= 0.46\textwidth]{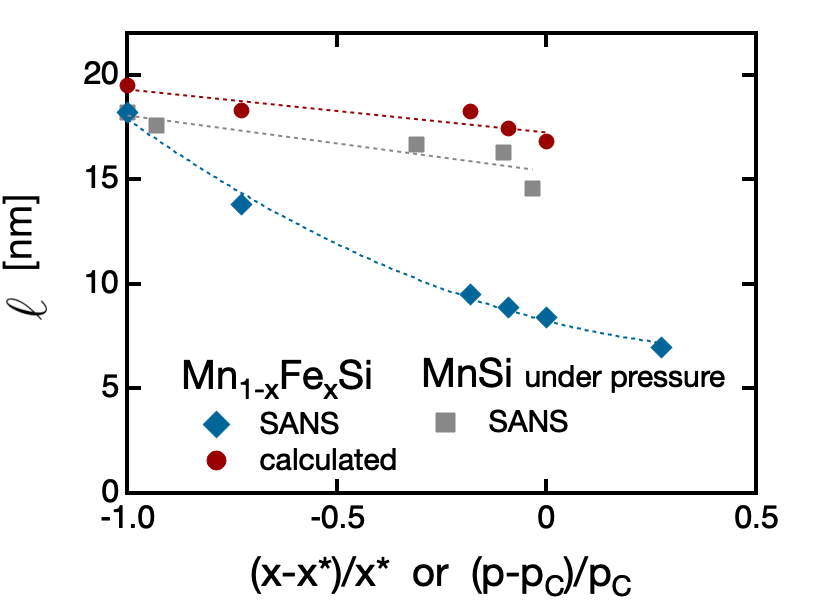}
\caption{Evolution of the helical pitch as a function of  Fe doping, in the case of Mn$_{1-x}$Fe$_{x}$Si, or as a function of pressure, in the case of MnSi. For the sake of comparison the absisca is either the relative doping $(x-x^*)/x^*$ for Mn$_{1-x}$Fe$_{x}$Si, or the relative pressure $(p-p_C)/p_C$ for MnSi. The red dots have been calculated for Mn$_{1-x}$Fe$_{x}$Si using the model of Eq. \ref{eq:model}.}
\label{pitch}
\end{center}
\end{figure}

By combining this experimental input we calculate the values of $\ell$  that are depicted in  Fig. \ref{pitch} and vary very little
with doping.  For the parent compound MnSi  the  relative difference  between the experimental and calculated values of $\ell$ does not exceed  6\%, which gives confidence to our approach. For the sake of comparison Fig. \ref{pitch}  also shows the pressure dependence of $\ell$ for MnSi (data from Fig. 2(e) of  \cite{Bannenberg2019pressure}), which is  weak and follows the trend of the calculated values.  This contrasts with the behaviour of   Mn$_{1-x}$Fe$_x$Si, where $\ell$ decreases rapidly with increasing Fe  doping. At x=0.11, the highest Fe concentration where it was still possible to determine both $M_0$ and $H_{C2}$, the experimental value of $\ell$ is half the one expected from the model of Eq. \ref{eq:model}.

An even more drastic reduction of the helical periodicity was recently  found in  MnSi$_{1-x}$Ge$_x$ \cite{TanigakiNanoLett, FujishiroNatCommun} and has been attributed to frustration  \cite{Mutter2019}.  The effect of frustration can be explained by the interplay between the DMI and higher-order derivatives terms in Eq. \ref{eq:model}. These were the original skyrmion stabilisation mechanism considered by Skyrme \cite{skyrme1961} and lead to an attracting skyrmion-skyrmion or soliton-soliton interaction  \cite{Leonov_Mostovoy_NatCom_2015}. 
The numerical simulations in  \cite{Mutter2019}  reveal a drastic decrease of $\ell$
in the presence of an  additional antiferromagnetic exchange $J_{AF}$ term in Eq. \ref{eq:model}. This reduction occurs even for weak frustration and, although it was not possible to derive an analytical expression of general validity,  the limit  $\ell \gg a  $, leads to \cite{Mutter2019}: 
\begin{equation}
{\cal \ell} \; \approx \;  2\pi a  \; \frac{\, J}{D} \left( 1 - 4 \; \frac{J_{AF} }{J }\, \right).
\label{eq:pitch2}
\end{equation}
Following this approximation for Mn$_{0.89}$Fe$_{0.11}$Si, where  $\ell/a \approx $ 18,  the 50\% pitch reduction leads to  $J_{AF}/J \approx1/8$. This relatively low ratio  points to weak frustration and is consistent with the  weak disorder seen by electron spin resonance \cite{demishev2013} and the exponential relaxation of the NSE spectra. 

Antiferromagnetic interactions may  result from  oscillatory  RKKY exchange between localised magnetic moments, as suggested by the analysis of the ordinary and anomalous Hall effects \cite{glushkov2015}.   These achiral  interactions increase with increasing $x$ and this effect would explain the destabilisation of the  helical periodicity at $x^*$. It was suggested that RKKY interactions arise from the modification of the  electron and hole concentrations by the substitution of Mn by Fe \cite{glushkov2015}.  Thus, similarly to MnSi under pressure, the modification of the electronic state  drives the destabilisation of the long  range helical order and of the sharp first order phase transition at $T_C$. However,  the specific microscopic mechanisms are different for the two systems as highlighted by the evolution of the helical pitch $\ell$ shown in Fig. \ref{pitch}. Frustration and disorder are important for Mn$_{1-x}$Fe$_{x}$Si in contrast to MnSi under pressure.

Frustration can stabilise spiral and skyrmion periodic states \cite{Leonov_Mostovoy_NatCom_2015, 2019_Science_frustr_SKL} but, unlike DMI, it does not impose the chirality i.e. it does not impose a sense of rotation.
This leads to  rich phase diagrams, with phases that are impossible in chiral magnets. The competition between frustration and DMI, however,  has only been touched upon in the literature \cite{Mutter2019, vonMalottki_Heinze_2019, Yuan_2017_Klaui}  and only in the low-temperature regime with a fixed magnetization modulus. 
In fact, the influence of frustration on the phase transition and the precursor states in chiral magnets is a most challenging and unresolved problem, which we approach  based  on general considerations. 

Near the ordering temperatures, the functional of Eq. \ref{eq:model} must be  supplemented by the Landau expansion in powers of the magnetization \cite{bak1980}: 
\begin{equation}
W_0= A\, (T-T_0)M^2+B \, M^4, 
\label{Landau}
\end{equation}
where $B>0$ and $T_0$ is the ferromagnetic ordering temperature, i.e.~the Curie temperature in   the absence  of  DMI. Under the influence of temperature and of an applied magnetic field, this term enables variations of the magnetization amplitude, which lead to sizeable effects in the vicinity of $T_C$, the transition temperature in the presence of DMI. In the precursor region of chiral magnets,  
the chiral twisting  is accompanied by  strong longitudinal modulations of the magnetization and its interplay with rotational modes \cite{rossler2006, Laliena_Campo_2016, leonov2018b, Shinozaki_Leonov_2017}. Thus, anomalous spin-textures can be expected such as a staggered half-skyrmion lattice \cite{rossler2006} - a close analogue of the square lattice of merons in frustrated magnets \cite{Kharkov_Maxim_2017}. 
Even in the absence of frustration these precursor modulated states   may have both senses of magnetization rotation. However, the modulus increases or decreases depending on whether the rotation adopts the correct  or the wrong rotational sense \cite{leonov2018b, Mukamel_1985} with the correct chirality eventually dominating close to $T_C$, as shown in Fig.\ref{intensity_chirality}.   This mechanism  induces fan-like oscillations of isolated skyrmions \cite{leonov2018b} and solitons \cite{Mukamel_1985} and leads to their attracting interaction with potentials containing a plethora of local minima at different mutual distances. 

	An increasing frustration  will  
amplify these precursor  modulations and will thus enhance and expand the precursor region,  
as it is indeed the case  for Mn$_x$Fe$_{1-x}$Si \cite{bannenberg2018mnfesisans, bannenberg2018mnfesisquid, bauer2010, grigoriev2009b, grigoriev2011}.


\section{Conclusion}

To conclude, our  results  shed light on  the evolution of helimagnetic correlations and fluctuations  in Mn$_{1-x}$Fe$_x$Si for $x<x_C$. 
On one hand, our observations  rule out both the spin liquid and the quantum fluctuations hypothesis for $x>x^*$, which cannot be  associated  with a quantum critical point.
On the other hand, our analysis indicates that, with increasing doping, frustration increases, which we attribute to increasing competition between  chiral DMI and achiral RKKY interactions. We argue that frustration  explains  the expansion of the precursor phase for $x<x_C$ and  the destabilisation of the long range heilmagnetic periodicity at $x^*$. 

More generally, both frustration and DMI can stabilise spiral and skyrmion periodic states, 
and the competition between the two
can lead to rich phase diagrams with new phases  and spin configurations, providing a fertile ground for future developments. 





\section{Acknowledgments}
\begin{acknowledgments}
C.P. thanks S.V. Grigoriev for useful discussions and for sharing unpublished experimental results. The experiments were performed at the Institut Laue Langevin, France, and the data are available at https://doi.ill.fr/10.5291/ILL-DATA.4-03-1729. The single crystals were aligned at the OrientExpress instrument with the support of Bachir Ouladdiaf. The work of L.J.B. and C.P.  has been financially supported by The Netherlands Organization for Scientific Research through Project No. 72 1.012 .102 (Larmor). CP acknowledges travel support from  JSPS KAKENHI Grant Number JP15H05882 (J-Physics).  ) 
\end{acknowledgments}


\begin{thebibliography}{70}%
\makeatletter
\providecommand \@ifxundefined [1]{%
 \@ifx{#1\undefined}
}%
\providecommand \@ifnum [1]{%
 \ifnum #1\expandafter \@firstoftwo
 \else \expandafter \@secondoftwo
 \fi
}%
\providecommand \@ifx [1]{%
 \ifx #1\expandafter \@firstoftwo
 \else \expandafter \@secondoftwo
 \fi
}%
\providecommand \natexlab [1]{#1}%
\providecommand \enquote  [1]{``#1''}%
\providecommand \bibnamefont  [1]{#1}%
\providecommand \bibfnamefont [1]{#1}%
\providecommand \citenamefont [1]{#1}%
\providecommand \href@noop [0]{\@secondoftwo}%
\providecommand \href [0]{\begingroup \@sanitize@url \@href}%
\providecommand \@href[1]{\@@startlink{#1}\@@href}%
\providecommand \@@href[1]{\endgroup#1\@@endlink}%
\providecommand \@sanitize@url [0]{\catcode `\\12\catcode `\$12\catcode
  `\&12\catcode `\#12\catcode `\^12\catcode `\_12\catcode `\%12\relax}%
\providecommand \@@startlink[1]{}%
\providecommand \@@endlink[0]{}%
\providecommand \url  [0]{\begingroup\@sanitize@url \@url }%
\providecommand \@url [1]{\endgroup\@href {#1}{\urlprefix }}%
\providecommand \urlprefix  [0]{URL }%
\providecommand \Eprint [0]{\href }%
\providecommand \doibase [0]{http://dx.doi.org/}%
\providecommand \selectlanguage [0]{\@gobble}%
\providecommand \bibinfo  [0]{\@secondoftwo}%
\providecommand \bibfield  [0]{\@secondoftwo}%
\providecommand \translation [1]{[#1]}%
\providecommand \BibitemOpen [0]{}%
\providecommand \bibitemStop [0]{}%
\providecommand \bibitemNoStop [0]{.\EOS\space}%
\providecommand \EOS [0]{\spacefactor3000\relax}%
\providecommand \BibitemShut  [1]{\csname bibitem#1\endcsname}%
\let\auto@bib@innerbib\@empty
\bibitem [{\citenamefont {M{\"u}hlbauer}\ \emph {et~al.}(2009)\citenamefont
  {M{\"u}hlbauer}, \citenamefont {Binz}, \citenamefont {Jonietz}, \citenamefont
  {Pfleiderer}, \citenamefont {Rosch}, \citenamefont {Neubauer}, \citenamefont
  {Georgii},\ and\ \citenamefont {B{\"o}ni}}]{muhlbauer2009}%
  \BibitemOpen
  \bibfield  {author} {\bibinfo {author} {\bibfnamefont {S.}~\bibnamefont
  {M{\"u}hlbauer}}, \bibinfo {author} {\bibfnamefont {B.}~\bibnamefont {Binz}},
  \bibinfo {author} {\bibfnamefont {F.}~\bibnamefont {Jonietz}}, \bibinfo
  {author} {\bibfnamefont {C.}~\bibnamefont {Pfleiderer}}, \bibinfo {author}
  {\bibfnamefont {A.}~\bibnamefont {Rosch}}, \bibinfo {author} {\bibfnamefont
  {A.}~\bibnamefont {Neubauer}}, \bibinfo {author} {\bibfnamefont
  {R.}~\bibnamefont {Georgii}}, \ and\ \bibinfo {author} {\bibfnamefont
  {P.}~\bibnamefont {B{\"o}ni}},\ }\href@noop {} {\bibfield  {journal}
  {\bibinfo  {journal} {Science}\ }\textbf {\bibinfo {volume} {323}},\ \bibinfo
  {pages} {915} (\bibinfo {year} {2009})}\BibitemShut {NoStop}%
\bibitem [{\citenamefont {Ishikawa}\ \emph {et~al.}(1977)\citenamefont
  {Ishikawa}, \citenamefont {Shirane}, \citenamefont {Tarvin},\ and\
  \citenamefont {Kohgi}}]{ishikawa1977}%
  \BibitemOpen
  \bibfield  {author} {\bibinfo {author} {\bibfnamefont {Y.}~\bibnamefont
  {Ishikawa}}, \bibinfo {author} {\bibfnamefont {G.}~\bibnamefont {Shirane}},
  \bibinfo {author} {\bibfnamefont {J.~A.}\ \bibnamefont {Tarvin}}, \ and\
  \bibinfo {author} {\bibfnamefont {M.}~\bibnamefont {Kohgi}},\ }\href@noop {}
  {\bibfield  {journal} {\bibinfo  {journal} {Phys. Rev. B}\ }\textbf {\bibinfo
  {volume} {16}},\ \bibinfo {pages} {4956} (\bibinfo {year}
  {1977})}\BibitemShut {NoStop}%
\bibitem [{\citenamefont {Thessieu}\ \emph {et~al.}(1995)\citenamefont
  {Thessieu}, \citenamefont {Flouquet}, \citenamefont {Lapertot}, \citenamefont
  {Stepanov},\ and\ \citenamefont {Jaccard}}]{Thessieu1995}%
  \BibitemOpen
  \bibfield  {author} {\bibinfo {author} {\bibfnamefont {C.}~\bibnamefont
  {Thessieu}}, \bibinfo {author} {\bibfnamefont {J.}~\bibnamefont {Flouquet}},
  \bibinfo {author} {\bibfnamefont {G.}~\bibnamefont {Lapertot}}, \bibinfo
  {author} {\bibfnamefont {A.~N.}\ \bibnamefont {Stepanov}}, \ and\ \bibinfo
  {author} {\bibfnamefont {D.}~\bibnamefont {Jaccard}},\ }\href@noop {}
  {\bibfield  {journal} {\bibinfo  {journal} {Solid State Communications}\
  }\textbf {\bibinfo {volume} {95}},\ \bibinfo {pages} {707} (\bibinfo {year}
  {1995})}\BibitemShut {NoStop}%
\bibitem [{\citenamefont {Corti}\ \emph {et~al.}(2007)\citenamefont {Corti},
  \citenamefont {Carbone}, \citenamefont {Filibian}, \citenamefont {Jarlborg},
  \citenamefont {Nugroho},\ and\ \citenamefont {Carretta}}]{Corti:2007kx}%
  \BibitemOpen
  \bibfield  {author} {\bibinfo {author} {\bibfnamefont {M.}~\bibnamefont
  {Corti}}, \bibinfo {author} {\bibfnamefont {F.}~\bibnamefont {Carbone}},
  \bibinfo {author} {\bibfnamefont {M.}~\bibnamefont {Filibian}}, \bibinfo
  {author} {\bibfnamefont {T.}~\bibnamefont {Jarlborg}}, \bibinfo {author}
  {\bibfnamefont {A.~A.}\ \bibnamefont {Nugroho}}, \ and\ \bibinfo {author}
  {\bibfnamefont {P.}~\bibnamefont {Carretta}},\ }\href@noop {} {\bibfield
  {journal} {\bibinfo  {journal} {Phys. Rev B}\ }\textbf {\bibinfo {volume}
  {75}},\ \bibinfo {pages} {115111} (\bibinfo {year} {2007})}\BibitemShut
  {NoStop}%
\bibitem [{\citenamefont {Demishev}\ \emph {et~al.}(2011)\citenamefont
  {Demishev}, \citenamefont {Semeno}, \citenamefont {Bogach}, \citenamefont
  {Glushkov}, \citenamefont {Sluchanko}, \citenamefont {Samarin},\ and\
  \citenamefont {Chernobrovkin}}]{demishev2011}%
  \BibitemOpen
  \bibfield  {author} {\bibinfo {author} {\bibfnamefont {S.~V.}\ \bibnamefont
  {Demishev}}, \bibinfo {author} {\bibfnamefont {A.~V.}\ \bibnamefont
  {Semeno}}, \bibinfo {author} {\bibfnamefont {A.~V.}\ \bibnamefont {Bogach}},
  \bibinfo {author} {\bibfnamefont {V.~V.}\ \bibnamefont {Glushkov}}, \bibinfo
  {author} {\bibfnamefont {N.~E.}\ \bibnamefont {Sluchanko}}, \bibinfo {author}
  {\bibfnamefont {N.~A.}\ \bibnamefont {Samarin}}, \ and\ \bibinfo {author}
  {\bibfnamefont {A.~L.}\ \bibnamefont {Chernobrovkin}},\ }\href@noop {}
  {\bibfield  {journal} {\bibinfo  {journal} {JETP Letters}\ }\textbf {\bibinfo
  {volume} {93}},\ \bibinfo {pages} {213} (\bibinfo {year} {2011})}\BibitemShut
  {NoStop}%
\bibitem [{\citenamefont {Yasuoka}\ \emph {et~al.}(2016)\citenamefont
  {Yasuoka}, \citenamefont {Motoya}, \citenamefont {Majumder}, \citenamefont
  {Physical},\ and\ \citenamefont {{2016}}}]{Yasuoka:2016hg}%
  \BibitemOpen
  \bibfield  {author} {\bibinfo {author} {\bibfnamefont {H.}~\bibnamefont
  {Yasuoka}}, \bibinfo {author} {\bibfnamefont {K.}~\bibnamefont {Motoya}},
  \bibinfo {author} {\bibfnamefont {M.}~\bibnamefont {Majumder}}, \bibinfo
  {author} {\bibfnamefont {S.~W. J. o.~t.}\ \bibnamefont {Physical}}, \ and\
  \bibinfo {author} {\bibnamefont {{2016}}},\ }\href@noop {} {\bibfield
  {journal} {\bibinfo  {journal} {journals.jps.jp}\ }\textbf {\bibinfo {volume}
  {85}},\ \bibinfo {pages} {073701} (\bibinfo {year} {2016})}\BibitemShut
  {NoStop}%
\bibitem [{\citenamefont {Bannenberg}\ \emph {et~al.}(2019)\citenamefont
  {Bannenberg}, \citenamefont {Sadykov}, \citenamefont {Dalgliesh},
  \citenamefont {Goodway}, \citenamefont {Schlagel}, \citenamefont {Lograsso},
  \citenamefont {Falus}, \citenamefont {Leli{\`e}vre-Berna}, \citenamefont
  {Leonov},\ and\ \citenamefont {Pappas}}]{Bannenberg2019pressure}%
  \BibitemOpen
  \bibfield  {author} {\bibinfo {author} {\bibfnamefont {L.~J.}\ \bibnamefont
  {Bannenberg}}, \bibinfo {author} {\bibfnamefont {R.}~\bibnamefont {Sadykov}},
  \bibinfo {author} {\bibfnamefont {R.~M.}\ \bibnamefont {Dalgliesh}}, \bibinfo
  {author} {\bibfnamefont {C.}~\bibnamefont {Goodway}}, \bibinfo {author}
  {\bibfnamefont {D.~L.}\ \bibnamefont {Schlagel}}, \bibinfo {author}
  {\bibfnamefont {T.~A.}\ \bibnamefont {Lograsso}}, \bibinfo {author}
  {\bibfnamefont {P.}~\bibnamefont {Falus}}, \bibinfo {author} {\bibfnamefont
  {E.}~\bibnamefont {Leli{\`e}vre-Berna}}, \bibinfo {author} {\bibfnamefont
  {A.~O.}\ \bibnamefont {Leonov}}, \ and\ \bibinfo {author} {\bibfnamefont
  {C.}~\bibnamefont {Pappas}},\ }\href@noop {} {\bibfield  {journal} {\bibinfo
  {journal} {Phys. Rev B}\ }\textbf {\bibinfo {volume} {100}},\ \bibinfo
  {pages} {054447} (\bibinfo {year} {2019})}\BibitemShut {NoStop}%
\bibitem [{\citenamefont {Yaouanc}\ \emph {et~al.}(2020)\citenamefont
  {Yaouanc}, \citenamefont {Dalmas~de R{\'e}otier}, \citenamefont {Roessli},
  \citenamefont {Maisuradze}, \citenamefont {Amato}, \citenamefont {Andreica},\
  and\ \citenamefont {Lapertot}}]{Yaouanc2020}%
  \BibitemOpen
  \bibfield  {author} {\bibinfo {author} {\bibfnamefont {A.}~\bibnamefont
  {Yaouanc}}, \bibinfo {author} {\bibfnamefont {P.}~\bibnamefont {Dalmas~de
  R{\'e}otier}}, \bibinfo {author} {\bibfnamefont {B.}~\bibnamefont {Roessli}},
  \bibinfo {author} {\bibfnamefont {A.}~\bibnamefont {Maisuradze}}, \bibinfo
  {author} {\bibfnamefont {A.}~\bibnamefont {Amato}}, \bibinfo {author}
  {\bibfnamefont {D.}~\bibnamefont {Andreica}}, \ and\ \bibinfo {author}
  {\bibfnamefont {G.}~\bibnamefont {Lapertot}},\ }\href@noop {} {\bibfield
  {journal} {\bibinfo  {journal} {Phys. Rev. Research}\ }\textbf {\bibinfo
  {volume} {2}},\ \bibinfo {pages} {013029} (\bibinfo {year}
  {2020})}\BibitemShut {NoStop}%
\bibitem [{\citenamefont {Pfleiderer}\ \emph {et~al.}(2001)\citenamefont
  {Pfleiderer}, \citenamefont {Julian},\ and\ \citenamefont
  {Lonzarich}}]{pfleiderer2001}%
  \BibitemOpen
  \bibfield  {author} {\bibinfo {author} {\bibfnamefont {C.}~\bibnamefont
  {Pfleiderer}}, \bibinfo {author} {\bibfnamefont {S.}~\bibnamefont {Julian}},
  \ and\ \bibinfo {author} {\bibfnamefont {G.~G.}\ \bibnamefont {Lonzarich}},\
  }\href@noop {} {\bibfield  {journal} {\bibinfo  {journal} {Nature}\ }\textbf
  {\bibinfo {volume} {414}},\ \bibinfo {pages} {427} (\bibinfo {year}
  {2001})}\BibitemShut {NoStop}%
\bibitem [{\citenamefont {Doiron-Leyraud}\ \emph {et~al.}(2003)\citenamefont
  {Doiron-Leyraud}, \citenamefont {Walker}, \citenamefont {Taillefer},
  \citenamefont {Steiner}, \citenamefont {Julian},\ and\ \citenamefont
  {Lonzarich}}]{doiron2003}%
  \BibitemOpen
  \bibfield  {author} {\bibinfo {author} {\bibfnamefont {N.}~\bibnamefont
  {Doiron-Leyraud}}, \bibinfo {author} {\bibfnamefont {I.~R.}\ \bibnamefont
  {Walker}}, \bibinfo {author} {\bibfnamefont {L.}~\bibnamefont {Taillefer}},
  \bibinfo {author} {\bibfnamefont {M.~J.}\ \bibnamefont {Steiner}}, \bibinfo
  {author} {\bibfnamefont {S.~R.}\ \bibnamefont {Julian}}, \ and\ \bibinfo
  {author} {\bibfnamefont {G.~G.}\ \bibnamefont {Lonzarich}},\ }\href@noop {}
  {\bibfield  {journal} {\bibinfo  {journal} {Nature}\ }\textbf {\bibinfo
  {volume} {425}},\ \bibinfo {pages} {595} (\bibinfo {year}
  {2003})}\BibitemShut {NoStop}%
\bibitem [{\citenamefont {Pfleiderer}\ \emph {et~al.}(2007)\citenamefont
  {Pfleiderer}, \citenamefont {B{\"o}ni}, \citenamefont {Keller}, \citenamefont
  {R{\"o}{\ss}ler},\ and\ \citenamefont {Rosch}}]{pfleiderer2007}%
  \BibitemOpen
  \bibfield  {author} {\bibinfo {author} {\bibfnamefont {C.}~\bibnamefont
  {Pfleiderer}}, \bibinfo {author} {\bibfnamefont {P.}~\bibnamefont
  {B{\"o}ni}}, \bibinfo {author} {\bibfnamefont {T.}~\bibnamefont {Keller}},
  \bibinfo {author} {\bibfnamefont {U.~K.}\ \bibnamefont {R{\"o}{\ss}ler}}, \
  and\ \bibinfo {author} {\bibfnamefont {A.}~\bibnamefont {Rosch}},\
  }\href@noop {} {\bibfield  {journal} {\bibinfo  {journal} {Science}\ }\textbf
  {\bibinfo {volume} {316}},\ \bibinfo {pages} {1871} (\bibinfo {year}
  {2007})}\BibitemShut {NoStop}%
\bibitem [{\citenamefont {Lee}\ \emph {et~al.}(2009)\citenamefont {Lee},
  \citenamefont {Kang}, \citenamefont {Onose}, \citenamefont {Tokura},\ and\
  \citenamefont {Ong}}]{lee2009}%
  \BibitemOpen
  \bibfield  {author} {\bibinfo {author} {\bibfnamefont {M.}~\bibnamefont
  {Lee}}, \bibinfo {author} {\bibfnamefont {W.}~\bibnamefont {Kang}}, \bibinfo
  {author} {\bibfnamefont {Y.}~\bibnamefont {Onose}}, \bibinfo {author}
  {\bibfnamefont {Y.}~\bibnamefont {Tokura}}, \ and\ \bibinfo {author}
  {\bibfnamefont {N.~P.}\ \bibnamefont {Ong}},\ }\href@noop {} {\bibfield
  {journal} {\bibinfo  {journal} {Phys. Rev. Lett.}\ }\textbf {\bibinfo
  {volume} {102}},\ \bibinfo {pages} {186601} (\bibinfo {year}
  {2009})}\BibitemShut {NoStop}%
\bibitem [{\citenamefont {Ritz}\ \emph {et~al.}(2013)\citenamefont {Ritz},
  \citenamefont {Halder}, \citenamefont {Wagner}, \citenamefont {Franz},
  \citenamefont {Bauer},\ and\ \citenamefont {Pfleiderer}}]{ritz2013}%
  \BibitemOpen
  \bibfield  {author} {\bibinfo {author} {\bibfnamefont {R.}~\bibnamefont
  {Ritz}}, \bibinfo {author} {\bibfnamefont {M.}~\bibnamefont {Halder}},
  \bibinfo {author} {\bibfnamefont {M.}~\bibnamefont {Wagner}}, \bibinfo
  {author} {\bibfnamefont {C.}~\bibnamefont {Franz}}, \bibinfo {author}
  {\bibfnamefont {A.}~\bibnamefont {Bauer}}, \ and\ \bibinfo {author}
  {\bibfnamefont {C.}~\bibnamefont {Pfleiderer}},\ }\href@noop {} {\bibfield
  {journal} {\bibinfo  {journal} {Nature}\ }\textbf {\bibinfo {volume} {497}},\
  \bibinfo {pages} {231} (\bibinfo {year} {2013})}\BibitemShut {NoStop}%
\bibitem [{\citenamefont {Kirkpatrick}\ and\ \citenamefont
  {Belitz}(2018)}]{Kirkpatrick2018}%
  \BibitemOpen
  \bibfield  {author} {\bibinfo {author} {\bibfnamefont {T.~R.}\ \bibnamefont
  {Kirkpatrick}}\ and\ \bibinfo {author} {\bibfnamefont {D.}~\bibnamefont
  {Belitz}},\ }\href@noop {} {\bibfield  {journal} {\bibinfo  {journal} {Phys.
  Rev B}\ }\textbf {\bibinfo {volume} {97}},\ \bibinfo {pages} {064411}
  (\bibinfo {year} {2018})}\BibitemShut {NoStop}%
\bibitem [{\citenamefont {Pfleiderer}\ \emph {et~al.}(1997)\citenamefont
  {Pfleiderer}, \citenamefont {McMullan}, \citenamefont {Julian},\ and\
  \citenamefont {Lonzarich}}]{pfleiderer1997}%
  \BibitemOpen
  \bibfield  {author} {\bibinfo {author} {\bibfnamefont {C.}~\bibnamefont
  {Pfleiderer}}, \bibinfo {author} {\bibfnamefont {G.~J.}\ \bibnamefont
  {McMullan}}, \bibinfo {author} {\bibfnamefont {S.~R.}\ \bibnamefont
  {Julian}}, \ and\ \bibinfo {author} {\bibfnamefont {G.~G.}\ \bibnamefont
  {Lonzarich}},\ }\href@noop {} {\bibfield  {journal} {\bibinfo  {journal}
  {Phys. Rev. B}\ }\textbf {\bibinfo {volume} {55}},\ \bibinfo {pages} {8330}
  (\bibinfo {year} {1997})}\BibitemShut {NoStop}%
\bibitem [{\citenamefont {Tewari}\ \emph {et~al.}(2006)\citenamefont {Tewari},
  \citenamefont {Belitz},\ and\ \citenamefont
  {Kirkpatrick}}]{BlueQuantumFog2006}%
  \BibitemOpen
  \bibfield  {author} {\bibinfo {author} {\bibfnamefont {S.}~\bibnamefont
  {Tewari}}, \bibinfo {author} {\bibfnamefont {D.}~\bibnamefont {Belitz}}, \
  and\ \bibinfo {author} {\bibfnamefont {T.~R.}\ \bibnamefont {Kirkpatrick}},\
  }\href@noop {} {\bibfield  {journal} {\bibinfo  {journal} {Phys. Rev. Lett.}\
  }\textbf {\bibinfo {volume} {96}},\ \bibinfo {pages} {047207} (\bibinfo
  {year} {2006})}\BibitemShut {NoStop}%
\bibitem [{\citenamefont {R{\"o}{\ss}ler}\ \emph {et~al.}(2006)\citenamefont
  {R{\"o}{\ss}ler}, \citenamefont {Bogdanov},\ and\ \citenamefont
  {Pfleiderer}}]{rossler2006}%
  \BibitemOpen
  \bibfield  {author} {\bibinfo {author} {\bibfnamefont {U.~K.}\ \bibnamefont
  {R{\"o}{\ss}ler}}, \bibinfo {author} {\bibfnamefont {A.~N.}\ \bibnamefont
  {Bogdanov}}, \ and\ \bibinfo {author} {\bibfnamefont {C.}~\bibnamefont
  {Pfleiderer}},\ }\href@noop {} {\bibfield  {journal} {\bibinfo  {journal}
  {Nature}\ }\textbf {\bibinfo {volume} {442}},\ \bibinfo {pages} {797}
  (\bibinfo {year} {2006})}\BibitemShut {NoStop}%
\bibitem [{\citenamefont {Hopkinson}\ and\ \citenamefont
  {Kee}(2009)}]{Hopkinson2009}%
  \BibitemOpen
  \bibfield  {author} {\bibinfo {author} {\bibfnamefont {J.~M.}\ \bibnamefont
  {Hopkinson}}\ and\ \bibinfo {author} {\bibfnamefont {H.-Y.}\ \bibnamefont
  {Kee}},\ }\href@noop {} {\bibfield  {journal} {\bibinfo  {journal} {Phys.
  Rev. B}\ }\textbf {\bibinfo {volume} {79}},\ \bibinfo {pages} {14421}
  (\bibinfo {year} {2009})}\BibitemShut {NoStop}%
\bibitem [{\citenamefont {Pfleiderer}\ \emph {et~al.}(2009)\citenamefont
  {Pfleiderer}, \citenamefont {Neubauer}, \citenamefont {Muhlbauer},
  \citenamefont {Jonietz}, \citenamefont {Janoschek}, \citenamefont {Legl},
  \citenamefont {Ritz}, \citenamefont {M{\"u}nzer}, \citenamefont {Franz},
  \citenamefont {Niklowitz}, \citenamefont {Keller}, \citenamefont {Georgii},
  \citenamefont {B{\"o}ni}, \citenamefont {Binz}, \citenamefont {Rosch},
  \citenamefont {R{\"o}{\ss}ler},\ and\ \citenamefont
  {Bogdanov}}]{Pfleiderer2009dr}%
  \BibitemOpen
  \bibfield  {author} {\bibinfo {author} {\bibfnamefont {C.}~\bibnamefont
  {Pfleiderer}}, \bibinfo {author} {\bibfnamefont {A.}~\bibnamefont
  {Neubauer}}, \bibinfo {author} {\bibfnamefont {S.}~\bibnamefont {Muhlbauer}},
  \bibinfo {author} {\bibfnamefont {F.}~\bibnamefont {Jonietz}}, \bibinfo
  {author} {\bibfnamefont {M.}~\bibnamefont {Janoschek}}, \bibinfo {author}
  {\bibfnamefont {S.}~\bibnamefont {Legl}}, \bibinfo {author} {\bibfnamefont
  {R.}~\bibnamefont {Ritz}}, \bibinfo {author} {\bibfnamefont {W.}~\bibnamefont
  {M{\"u}nzer}}, \bibinfo {author} {\bibfnamefont {C.}~\bibnamefont {Franz}},
  \bibinfo {author} {\bibfnamefont {P.~G.}\ \bibnamefont {Niklowitz}}, \bibinfo
  {author} {\bibfnamefont {T.}~\bibnamefont {Keller}}, \bibinfo {author}
  {\bibfnamefont {R.}~\bibnamefont {Georgii}}, \bibinfo {author} {\bibfnamefont
  {P.}~\bibnamefont {B{\"o}ni}}, \bibinfo {author} {\bibfnamefont
  {B.}~\bibnamefont {Binz}}, \bibinfo {author} {\bibfnamefont {A.}~\bibnamefont
  {Rosch}}, \bibinfo {author} {\bibfnamefont {U.~K.}\ \bibnamefont
  {R{\"o}{\ss}ler}}, \ and\ \bibinfo {author} {\bibfnamefont {A.~N.}\
  \bibnamefont {Bogdanov}},\ }\href@noop {} {\bibfield  {journal} {\bibinfo
  {journal} {Journal of Physics-Condensed Matter}\ }\textbf {\bibinfo {volume}
  {21}},\ \bibinfo {pages} {164215} (\bibinfo {year} {2009})}\BibitemShut
  {NoStop}%
\bibitem [{\citenamefont {Kirkpatrick}\ and\ \citenamefont
  {Belitz}(2010)}]{KirkpatrickPRL2010}%
  \BibitemOpen
  \bibfield  {author} {\bibinfo {author} {\bibfnamefont {T.~R.}\ \bibnamefont
  {Kirkpatrick}}\ and\ \bibinfo {author} {\bibfnamefont {D.}~\bibnamefont
  {Belitz}},\ }\href@noop {} {\bibfield  {journal} {\bibinfo  {journal} {Phys.
  Rev. Lett.}\ }\textbf {\bibinfo {volume} {104}},\ \bibinfo {pages} {256404}
  (\bibinfo {year} {2010})}\BibitemShut {NoStop}%
\bibitem [{\citenamefont {Ho}\ \emph {et~al.}(2010)\citenamefont {Ho},
  \citenamefont {Kirkpatrick}, \citenamefont {Sang},\ and\ \citenamefont
  {Belitz}}]{KirkpatrickPRB2010}%
  \BibitemOpen
  \bibfield  {author} {\bibinfo {author} {\bibfnamefont {K.-Y.}\ \bibnamefont
  {Ho}}, \bibinfo {author} {\bibfnamefont {T.~R.}\ \bibnamefont {Kirkpatrick}},
  \bibinfo {author} {\bibfnamefont {Y.}~\bibnamefont {Sang}}, \ and\ \bibinfo
  {author} {\bibfnamefont {D.}~\bibnamefont {Belitz}},\ }\href@noop {}
  {\bibfield  {journal} {\bibinfo  {journal} {Phys. Rev. B}\ }\textbf {\bibinfo
  {volume} {82}},\ \bibinfo {pages} {134427} (\bibinfo {year}
  {2010})}\BibitemShut {NoStop}%
\bibitem [{\citenamefont {Kr{\"u}ger}\ \emph {et~al.}(2012)\citenamefont
  {Kr{\"u}ger}, \citenamefont {Karahasanovic},\ and\ \citenamefont
  {Green}}]{Kruger2012}%
  \BibitemOpen
  \bibfield  {author} {\bibinfo {author} {\bibfnamefont {F.}~\bibnamefont
  {Kr{\"u}ger}}, \bibinfo {author} {\bibfnamefont {U.}~\bibnamefont
  {Karahasanovic}}, \ and\ \bibinfo {author} {\bibfnamefont {A.~G.}\
  \bibnamefont {Green}},\ }\href@noop {} {\bibfield  {journal} {\bibinfo
  {journal} {Phys. Rev. Lett.}\ }\textbf {\bibinfo {volume} {108}},\ \bibinfo
  {pages} {067003} (\bibinfo {year} {2012})}\BibitemShut {NoStop}%
\bibitem [{\citenamefont {Povzner}\ \emph {et~al.}(2018)\citenamefont
  {Povzner}, \citenamefont {Volkov},\ and\ \citenamefont
  {Nogovitsyna}}]{Povzner2018}%
  \BibitemOpen
  \bibfield  {author} {\bibinfo {author} {\bibfnamefont {A.~A.}\ \bibnamefont
  {Povzner}}, \bibinfo {author} {\bibfnamefont {A.~G.}\ \bibnamefont {Volkov}},
  \ and\ \bibinfo {author} {\bibfnamefont {T.~A.}\ \bibnamefont
  {Nogovitsyna}},\ }\href@noop {} {\bibfield  {journal} {\bibinfo  {journal}
  {Physica B: Physics of Condensed Matter}\ }\textbf {\bibinfo {volume}
  {536}},\ \bibinfo {pages} {408} (\bibinfo {year} {2018})}\BibitemShut
  {NoStop}%
\bibitem [{\citenamefont {Pfleiderer}\ \emph {et~al.}(2004)\citenamefont
  {Pfleiderer}, \citenamefont {Reznik}, \citenamefont {Pintschovius},
  \citenamefont {v~Lohneysen} \emph {et~al.}}]{pfleiderer2004}%
  \BibitemOpen
  \bibfield  {author} {\bibinfo {author} {\bibfnamefont {C.}~\bibnamefont
  {Pfleiderer}}, \bibinfo {author} {\bibfnamefont {D.}~\bibnamefont {Reznik}},
  \bibinfo {author} {\bibfnamefont {L.}~\bibnamefont {Pintschovius}}, \bibinfo
  {author} {\bibfnamefont {H.}~\bibnamefont {v~Lohneysen}},  \emph {et~al.},\
  }\href@noop {} {\bibfield  {journal} {\bibinfo  {journal} {Nature}\ }\textbf
  {\bibinfo {volume} {427}},\ \bibinfo {pages} {227} (\bibinfo {year}
  {2004})}\BibitemShut {NoStop}%
\bibitem [{\citenamefont {Pintschovius}\ \emph {et~al.}(2004)\citenamefont
  {Pintschovius}, \citenamefont {Reznik}, \citenamefont {Pfleiderer},\ and\
  \citenamefont {von L{\"o}hneysen}}]{pintschovius2004}%
  \BibitemOpen
  \bibfield  {author} {\bibinfo {author} {\bibfnamefont {L.}~\bibnamefont
  {Pintschovius}}, \bibinfo {author} {\bibfnamefont {D.}~\bibnamefont
  {Reznik}}, \bibinfo {author} {\bibfnamefont {C.}~\bibnamefont {Pfleiderer}},
  \ and\ \bibinfo {author} {\bibfnamefont {H.}~\bibnamefont {von
  L{\"o}hneysen}},\ }\href@noop {} {\bibfield  {journal} {\bibinfo  {journal}
  {Pramana}\ }\textbf {\bibinfo {volume} {63}},\ \bibinfo {pages} {117}
  (\bibinfo {year} {2004})}\BibitemShut {NoStop}%
\bibitem [{\citenamefont {Binz}\ \emph {et~al.}(2006)\citenamefont {Binz},
  \citenamefont {Vishwanath},\ and\ \citenamefont {Aji}}]{Binz2006}%
  \BibitemOpen
  \bibfield  {author} {\bibinfo {author} {\bibfnamefont {B.}~\bibnamefont
  {Binz}}, \bibinfo {author} {\bibfnamefont {A.}~\bibnamefont {Vishwanath}}, \
  and\ \bibinfo {author} {\bibfnamefont {V.}~\bibnamefont {Aji}},\ }\href@noop
  {} {\bibfield  {journal} {\bibinfo  {journal} {Phys. Rev. Lett.}\ }\textbf
  {\bibinfo {volume} {96}},\ \bibinfo {pages} {207202} (\bibinfo {year}
  {2006})}\BibitemShut {NoStop}%
\bibitem [{\citenamefont {Fischer}\ \emph {et~al.}(2008)\citenamefont
  {Fischer}, \citenamefont {Shah},\ and\ \citenamefont {Rosch}}]{Fischer2008}%
  \BibitemOpen
  \bibfield  {author} {\bibinfo {author} {\bibfnamefont {I.}~\bibnamefont
  {Fischer}}, \bibinfo {author} {\bibfnamefont {N.}~\bibnamefont {Shah}}, \
  and\ \bibinfo {author} {\bibfnamefont {A.}~\bibnamefont {Rosch}},\
  }\href@noop {} {\bibfield  {journal} {\bibinfo  {journal} {Phys. Rev B}\
  }\textbf {\bibinfo {volume} {77}},\ \bibinfo {pages} {024415} (\bibinfo
  {year} {2008})}\BibitemShut {NoStop}%
\bibitem [{\citenamefont {Bannenberg}\ \emph
  {et~al.}(2018{\natexlab{a}})\citenamefont {Bannenberg}, \citenamefont
  {Dalgliesh}, \citenamefont {Wolf}, \citenamefont {Weber},\ and\ \citenamefont
  {Pappas}}]{bannenberg2018mnfesisans}%
  \BibitemOpen
  \bibfield  {author} {\bibinfo {author} {\bibfnamefont {L.~J.}\ \bibnamefont
  {Bannenberg}}, \bibinfo {author} {\bibfnamefont {R.~M.}\ \bibnamefont
  {Dalgliesh}}, \bibinfo {author} {\bibfnamefont {T.}~\bibnamefont {Wolf}},
  \bibinfo {author} {\bibfnamefont {F.}~\bibnamefont {Weber}}, \ and\ \bibinfo
  {author} {\bibfnamefont {C.}~\bibnamefont {Pappas}},\ }\href@noop {}
  {\bibfield  {journal} {\bibinfo  {journal} {Phys. Rev. B}\ }\textbf {\bibinfo
  {volume} {98}},\ \bibinfo {pages} {184431} (\bibinfo {year}
  {2018}{\natexlab{a}})}\BibitemShut {NoStop}%
\bibitem [{\citenamefont {Bannenberg}\ \emph
  {et~al.}(2018{\natexlab{b}})\citenamefont {Bannenberg}, \citenamefont
  {Weber}, \citenamefont {Lefering}, \citenamefont {Wolf},\ and\ \citenamefont
  {Pappas}}]{bannenberg2018mnfesisquid}%
  \BibitemOpen
  \bibfield  {author} {\bibinfo {author} {\bibfnamefont {L.~J.}\ \bibnamefont
  {Bannenberg}}, \bibinfo {author} {\bibfnamefont {F.}~\bibnamefont {Weber}},
  \bibinfo {author} {\bibfnamefont {A.~J.~E.}\ \bibnamefont {Lefering}},
  \bibinfo {author} {\bibfnamefont {T.}~\bibnamefont {Wolf}}, \ and\ \bibinfo
  {author} {\bibfnamefont {C.}~\bibnamefont {Pappas}},\ }\href@noop {}
  {\bibfield  {journal} {\bibinfo  {journal} {Phys. Rev. B}\ }\textbf {\bibinfo
  {volume} {98}},\ \bibinfo {pages} {184430} (\bibinfo {year}
  {2018}{\natexlab{b}})}\BibitemShut {NoStop}%
\bibitem [{\citenamefont {Bauer}\ \emph {et~al.}(2010)\citenamefont {Bauer},
  \citenamefont {Neubauer}, \citenamefont {Franz}, \citenamefont {M{\"u}nzer},
  \citenamefont {Garst},\ and\ \citenamefont {Pfleiderer}}]{bauer2010}%
  \BibitemOpen
  \bibfield  {author} {\bibinfo {author} {\bibfnamefont {A.}~\bibnamefont
  {Bauer}}, \bibinfo {author} {\bibfnamefont {A.}~\bibnamefont {Neubauer}},
  \bibinfo {author} {\bibfnamefont {C.}~\bibnamefont {Franz}}, \bibinfo
  {author} {\bibfnamefont {W.}~\bibnamefont {M{\"u}nzer}}, \bibinfo {author}
  {\bibfnamefont {M.}~\bibnamefont {Garst}}, \ and\ \bibinfo {author}
  {\bibfnamefont {C.}~\bibnamefont {Pfleiderer}},\ }\href@noop {} {\bibfield
  {journal} {\bibinfo  {journal} {Phys. Rev. B}\ }\textbf {\bibinfo {volume}
  {82}},\ \bibinfo {pages} {064404} (\bibinfo {year} {2010})}\BibitemShut
  {NoStop}%
\bibitem [{\citenamefont {Grigoriev}\ \emph {et~al.}(2009)\citenamefont
  {Grigoriev}, \citenamefont {Dyadkin}, \citenamefont {Moskvin}, \citenamefont
  {Lamago}, \citenamefont {Wolf}, \citenamefont {Eckerlebe},\ and\
  \citenamefont {Maleyev}}]{grigoriev2009b}%
  \BibitemOpen
  \bibfield  {author} {\bibinfo {author} {\bibfnamefont {S.~V.}\ \bibnamefont
  {Grigoriev}}, \bibinfo {author} {\bibfnamefont {V.~A.}\ \bibnamefont
  {Dyadkin}}, \bibinfo {author} {\bibfnamefont {E.~V.}\ \bibnamefont
  {Moskvin}}, \bibinfo {author} {\bibfnamefont {D.}~\bibnamefont {Lamago}},
  \bibinfo {author} {\bibfnamefont {T.}~\bibnamefont {Wolf}}, \bibinfo {author}
  {\bibfnamefont {H.}~\bibnamefont {Eckerlebe}}, \ and\ \bibinfo {author}
  {\bibfnamefont {S.~V.}\ \bibnamefont {Maleyev}},\ }\href@noop {} {\bibfield
  {journal} {\bibinfo  {journal} {Phys. Rev. B}\ }\textbf {\bibinfo {volume}
  {79}},\ \bibinfo {pages} {144417} (\bibinfo {year} {2009})}\BibitemShut
  {NoStop}%
\bibitem [{\citenamefont {Grigoriev}\ \emph {et~al.}(2011)\citenamefont
  {Grigoriev}, \citenamefont {Moskvin}, \citenamefont {Dyadkin}, \citenamefont
  {Lamago}, \citenamefont {Wolf}, \citenamefont {Eckerlebe},\ and\
  \citenamefont {Maleyev}}]{grigoriev2011}%
  \BibitemOpen
  \bibfield  {author} {\bibinfo {author} {\bibfnamefont {S.~V.}\ \bibnamefont
  {Grigoriev}}, \bibinfo {author} {\bibfnamefont {E.~V.}\ \bibnamefont
  {Moskvin}}, \bibinfo {author} {\bibfnamefont {V.~A.}\ \bibnamefont
  {Dyadkin}}, \bibinfo {author} {\bibfnamefont {D.}~\bibnamefont {Lamago}},
  \bibinfo {author} {\bibfnamefont {T.}~\bibnamefont {Wolf}}, \bibinfo {author}
  {\bibfnamefont {H.}~\bibnamefont {Eckerlebe}}, \ and\ \bibinfo {author}
  {\bibfnamefont {S.~V.}\ \bibnamefont {Maleyev}},\ }\href@noop {} {\bibfield
  {journal} {\bibinfo  {journal} {Phys. Rev. B}\ }\textbf {\bibinfo {volume}
  {83}},\ \bibinfo {pages} {224411} (\bibinfo {year} {2011})}\BibitemShut
  {NoStop}%
\bibitem [{\citenamefont {Franz}\ \emph {et~al.}(2014)\citenamefont {Franz},
  \citenamefont {Freimuth}, \citenamefont {Bauer}, \citenamefont {Ritz},
  \citenamefont {Schnarr}, \citenamefont {Duvinage}, \citenamefont {Adams},
  \citenamefont {Bl{\"u}gel}, \citenamefont {Rosch}, \citenamefont {Mokrousov}
  \emph {et~al.}}]{franz2014}%
  \BibitemOpen
  \bibfield  {author} {\bibinfo {author} {\bibfnamefont {C.}~\bibnamefont
  {Franz}}, \bibinfo {author} {\bibfnamefont {F.}~\bibnamefont {Freimuth}},
  \bibinfo {author} {\bibfnamefont {A.}~\bibnamefont {Bauer}}, \bibinfo
  {author} {\bibfnamefont {R.}~\bibnamefont {Ritz}}, \bibinfo {author}
  {\bibfnamefont {C.}~\bibnamefont {Schnarr}}, \bibinfo {author} {\bibfnamefont
  {C.}~\bibnamefont {Duvinage}}, \bibinfo {author} {\bibfnamefont
  {T.}~\bibnamefont {Adams}}, \bibinfo {author} {\bibfnamefont
  {S.}~\bibnamefont {Bl{\"u}gel}}, \bibinfo {author} {\bibfnamefont
  {A.}~\bibnamefont {Rosch}}, \bibinfo {author} {\bibfnamefont
  {Y.}~\bibnamefont {Mokrousov}},  \emph {et~al.},\ }\href@noop {} {\bibfield
  {journal} {\bibinfo  {journal} {Phys. Rev. Lett.}\ }\textbf {\bibinfo
  {volume} {112}},\ \bibinfo {pages} {186601} (\bibinfo {year}
  {2014})}\BibitemShut {NoStop}%
\bibitem [{\citenamefont {Petrova}\ \emph {et~al.}(2019)\citenamefont
  {Petrova}, \citenamefont {Gavrilkin}, \citenamefont {Menzel},\ and\
  \citenamefont {Stishov}}]{Petrova_2019_MnFeSi}%
  \BibitemOpen
  \bibfield  {author} {\bibinfo {author} {\bibfnamefont {A.~E.}\ \bibnamefont
  {Petrova}}, \bibinfo {author} {\bibfnamefont {S.~Y.}\ \bibnamefont
  {Gavrilkin}}, \bibinfo {author} {\bibfnamefont {D.}~\bibnamefont {Menzel}}, \
  and\ \bibinfo {author} {\bibfnamefont {S.~M.}\ \bibnamefont {Stishov}},\
  }\href@noop {} {\bibfield  {journal} {\bibinfo  {journal} {Phys. Rev B}\
  }\textbf {\bibinfo {volume} {100}},\ \bibinfo {pages} {094403} (\bibinfo
  {year} {2019})}\BibitemShut {NoStop}%
\bibitem [{\citenamefont {Demishev}\ \emph {et~al.}(2014)\citenamefont
  {Demishev}, \citenamefont {Samarin}, \citenamefont {Glushkov}, \citenamefont
  {Gilmanov}, \citenamefont {Lobanova}, \citenamefont {Samarin}, \citenamefont
  {Semeno}, \citenamefont {Sluchanko}, \citenamefont {Chubova}, \citenamefont
  {Dyadkin},\ and\ \citenamefont {Grigoriev}}]{demishev2014}%
  \BibitemOpen
  \bibfield  {author} {\bibinfo {author} {\bibfnamefont {S.~V.}\ \bibnamefont
  {Demishev}}, \bibinfo {author} {\bibfnamefont {A.~N.}\ \bibnamefont
  {Samarin}}, \bibinfo {author} {\bibfnamefont {V.~V.}\ \bibnamefont
  {Glushkov}}, \bibinfo {author} {\bibfnamefont {M.~I.}\ \bibnamefont
  {Gilmanov}}, \bibinfo {author} {\bibfnamefont {I.~I.}\ \bibnamefont
  {Lobanova}}, \bibinfo {author} {\bibfnamefont {N.~A.}\ \bibnamefont
  {Samarin}}, \bibinfo {author} {\bibfnamefont {A.~V.}\ \bibnamefont {Semeno}},
  \bibinfo {author} {\bibfnamefont {N.~E.}\ \bibnamefont {Sluchanko}}, \bibinfo
  {author} {\bibfnamefont {N.~M.}\ \bibnamefont {Chubova}}, \bibinfo {author}
  {\bibfnamefont {V.~A.}\ \bibnamefont {Dyadkin}}, \ and\ \bibinfo {author}
  {\bibfnamefont {S.~V.}\ \bibnamefont {Grigoriev}},\ }\href@noop {} {\bibfield
   {journal} {\bibinfo  {journal} {JETP Letters}\ }\textbf {\bibinfo {volume}
  {100}},\ \bibinfo {pages} {28} (\bibinfo {year} {2014})}\BibitemShut
  {NoStop}%
\bibitem [{\citenamefont {Demishev}\ \emph {et~al.}(2016)\citenamefont
  {Demishev}, \citenamefont {Samarin}, \citenamefont {Huang}, \citenamefont
  {Glushkov}, \citenamefont {Lobanova}, \citenamefont {Sluchanko},
  \citenamefont {Chubova}, \citenamefont {Dyadkin}, \citenamefont {Grigoriev},
  \citenamefont {Kagan} \emph {et~al.}}]{demishev2016a}%
  \BibitemOpen
  \bibfield  {author} {\bibinfo {author} {\bibfnamefont {S.~V.}\ \bibnamefont
  {Demishev}}, \bibinfo {author} {\bibfnamefont {A.~N.}\ \bibnamefont
  {Samarin}}, \bibinfo {author} {\bibfnamefont {J.}~\bibnamefont {Huang}},
  \bibinfo {author} {\bibfnamefont {V.~V.}\ \bibnamefont {Glushkov}}, \bibinfo
  {author} {\bibfnamefont {I.~I.}\ \bibnamefont {Lobanova}}, \bibinfo {author}
  {\bibfnamefont {N.~E.}\ \bibnamefont {Sluchanko}}, \bibinfo {author}
  {\bibfnamefont {N.~M.}\ \bibnamefont {Chubova}}, \bibinfo {author}
  {\bibfnamefont {V.~A.}\ \bibnamefont {Dyadkin}}, \bibinfo {author}
  {\bibfnamefont {S.~V.}\ \bibnamefont {Grigoriev}}, \bibinfo {author}
  {\bibfnamefont {M.~Y.}\ \bibnamefont {Kagan}},  \emph {et~al.},\ }\href@noop
  {} {\bibfield  {journal} {\bibinfo  {journal} {JETP letters}\ }\textbf
  {\bibinfo {volume} {104}},\ \bibinfo {pages} {116} (\bibinfo {year}
  {2016})}\BibitemShut {NoStop}%
\bibitem [{\citenamefont {Kindervater}\ \emph {et~al.}(2020)\citenamefont
  {Kindervater}, \citenamefont {Adams}, \citenamefont {Bauer}, \citenamefont
  {Haslbeck}, \citenamefont {Chacon}, \citenamefont {M\"uhlbauer},
  \citenamefont {Jonietz}, \citenamefont {Neubauer}, \citenamefont {Gasser},
  \citenamefont {Nagy}, \citenamefont {Martin}, \citenamefont {H\"au\ss{}ler},
  \citenamefont {Georgii}, \citenamefont {Garst},\ and\ \citenamefont
  {Pfleiderer}}]{KindervaterPRB2020}%
  \BibitemOpen
  \bibfield  {author} {\bibinfo {author} {\bibfnamefont {J.}~\bibnamefont
  {Kindervater}}, \bibinfo {author} {\bibfnamefont {T.}~\bibnamefont {Adams}},
  \bibinfo {author} {\bibfnamefont {A.}~\bibnamefont {Bauer}}, \bibinfo
  {author} {\bibfnamefont {F.~X.}\ \bibnamefont {Haslbeck}}, \bibinfo {author}
  {\bibfnamefont {A.}~\bibnamefont {Chacon}}, \bibinfo {author} {\bibfnamefont
  {S.}~\bibnamefont {M\"uhlbauer}}, \bibinfo {author} {\bibfnamefont
  {F.}~\bibnamefont {Jonietz}}, \bibinfo {author} {\bibfnamefont
  {A.}~\bibnamefont {Neubauer}}, \bibinfo {author} {\bibfnamefont
  {U.}~\bibnamefont {Gasser}}, \bibinfo {author} {\bibfnamefont
  {G.}~\bibnamefont {Nagy}}, \bibinfo {author} {\bibfnamefont {N.}~\bibnamefont
  {Martin}}, \bibinfo {author} {\bibfnamefont {W.}~\bibnamefont
  {H\"au\ss{}ler}}, \bibinfo {author} {\bibfnamefont {R.}~\bibnamefont
  {Georgii}}, \bibinfo {author} {\bibfnamefont {M.}~\bibnamefont {Garst}}, \
  and\ \bibinfo {author} {\bibfnamefont {C.}~\bibnamefont {Pfleiderer}},\
  }\href {\doibase 10.1103/PhysRevB.101.104406} {\bibfield  {journal} {\bibinfo
   {journal} {Phys. Rev. B}\ }\textbf {\bibinfo {volume} {101}},\ \bibinfo
  {pages} {104406} (\bibinfo {year} {2020})}\BibitemShut {NoStop}%
\bibitem [{\citenamefont {Blume}(1963)}]{Blume}%
  \BibitemOpen
  \bibfield  {author} {\bibinfo {author} {\bibfnamefont {M.}~\bibnamefont
  {Blume}},\ }\href@noop {} {\bibfield  {journal} {\bibinfo  {journal} {Phys.
  Rev.}\ }\textbf {\bibinfo {volume} {130}},\ \bibinfo {pages} {1670} (\bibinfo
  {year} {1963})}\BibitemShut {NoStop}%
\bibitem [{\citenamefont {Maleyev}\ \emph {et~al.}(1963)\citenamefont
  {Maleyev}, \citenamefont {Baryakhtar},\ and\ \citenamefont
  {Suris}}]{Maleyev2}%
  \BibitemOpen
  \bibfield  {author} {\bibinfo {author} {\bibfnamefont {S.~V.}\ \bibnamefont
  {Maleyev}}, \bibinfo {author} {\bibfnamefont {V.~G.}\ \bibnamefont
  {Baryakhtar}}, \ and\ \bibinfo {author} {\bibfnamefont {P.~A.}\ \bibnamefont
  {Suris}},\ }\href@noop {} {\bibfield  {journal} {\bibinfo  {journal} {Sov.
  Phys. Solid State}\ }\textbf {\bibinfo {volume} {4}},\ \bibinfo {pages}
  {2533} (\bibinfo {year} {1963})}\BibitemShut {NoStop}%
\bibitem [{\citenamefont {Murani}\ and\ \citenamefont
  {Mezei}(1980)}]{Murani:1980vg}%
  \BibitemOpen
  \bibfield  {author} {\bibinfo {author} {\bibfnamefont {A.~P.}\ \bibnamefont
  {Murani}}\ and\ \bibinfo {author} {\bibfnamefont {F.}~\bibnamefont {Mezei}},\
  }in\ \href@noop {} {\emph {\bibinfo {booktitle} {Neutron Spin Echo}}}\
  (\bibinfo  {publisher} {Lecture Notes in Physics, vol 128. Springer, Berlin,
  Heidelberg},\ \bibinfo {year} {1980})\ p.\ \bibinfo {pages} {104}\BibitemShut
  {NoStop}%
\bibitem [{\citenamefont {Pappas}\ \emph {et~al.}(2006)\citenamefont {Pappas},
  \citenamefont {Ehlers},\ and\ \citenamefont {Mezei}}]{PAPPAS2006521}%
  \BibitemOpen
  \bibfield  {author} {\bibinfo {author} {\bibfnamefont {C.}~\bibnamefont
  {Pappas}}, \bibinfo {author} {\bibfnamefont {G.}~\bibnamefont {Ehlers}}, \
  and\ \bibinfo {author} {\bibfnamefont {F.}~\bibnamefont {Mezei}},\ }in\ \href
  {\doibase https://doi.org/10.1016/B978-044451050-1/50012-4} {\emph {\bibinfo
  {booktitle} {Neutron Scattering from Magnetic Materials}}},\ \bibinfo
  {editor} {edited by\ \bibinfo {editor} {\bibfnamefont {T.}~\bibnamefont
  {Chatterji}}}\ (\bibinfo  {publisher} {Elsevier},\ \bibinfo {year} {2006})\
  p.\ \bibinfo {pages} {521}\BibitemShut {NoStop}%
\bibitem [{\citenamefont {Pappas}\ \emph {et~al.}(2009)\citenamefont {Pappas},
  \citenamefont {Lelievre-Berna}, \citenamefont {Falus}, \citenamefont
  {Bentley}, \citenamefont {Moskvin}, \citenamefont {Grigoriev}, \citenamefont
  {Fouquet},\ and\ \citenamefont {Farago}}]{pappas2009}%
  \BibitemOpen
  \bibfield  {author} {\bibinfo {author} {\bibfnamefont {C.}~\bibnamefont
  {Pappas}}, \bibinfo {author} {\bibfnamefont {E.}~\bibnamefont
  {Lelievre-Berna}}, \bibinfo {author} {\bibfnamefont {P.}~\bibnamefont
  {Falus}}, \bibinfo {author} {\bibfnamefont {P.~M.}\ \bibnamefont {Bentley}},
  \bibinfo {author} {\bibfnamefont {E.}~\bibnamefont {Moskvin}}, \bibinfo
  {author} {\bibfnamefont {S.}~\bibnamefont {Grigoriev}}, \bibinfo {author}
  {\bibfnamefont {P.}~\bibnamefont {Fouquet}}, \ and\ \bibinfo {author}
  {\bibfnamefont {B.}~\bibnamefont {Farago}},\ }\href@noop {} {\bibfield
  {journal} {\bibinfo  {journal} {Phys. Rev. Lett.}\ }\textbf {\bibinfo
  {volume} {102}},\ \bibinfo {pages} {197202} (\bibinfo {year}
  {2009})}\BibitemShut {NoStop}%
\bibitem [{\citenamefont {Pappas}\ \emph {et~al.}(2011)\citenamefont {Pappas},
  \citenamefont {Lelievre-Berna}, \citenamefont {Bentley}, \citenamefont
  {Falus}, \citenamefont {Fouquet},\ and\ \citenamefont {Farago}}]{pappas2011}%
  \BibitemOpen
  \bibfield  {author} {\bibinfo {author} {\bibfnamefont {C.}~\bibnamefont
  {Pappas}}, \bibinfo {author} {\bibfnamefont {E.}~\bibnamefont
  {Lelievre-Berna}}, \bibinfo {author} {\bibfnamefont {P.}~\bibnamefont
  {Bentley}}, \bibinfo {author} {\bibfnamefont {P.}~\bibnamefont {Falus}},
  \bibinfo {author} {\bibfnamefont {P.}~\bibnamefont {Fouquet}}, \ and\
  \bibinfo {author} {\bibfnamefont {B.}~\bibnamefont {Farago}},\ }\href@noop {}
  {\bibfield  {journal} {\bibinfo  {journal} {Phys. Rev. B}\ }\textbf {\bibinfo
  {volume} {83}},\ \bibinfo {pages} {224405} (\bibinfo {year}
  {2011})}\BibitemShut {NoStop}%
\bibitem [{\citenamefont {Janoschek}\ \emph {et~al.}(2013)\citenamefont
  {Janoschek}, \citenamefont {Garst}, \citenamefont {Bauer}, \citenamefont
  {Krautscheid}, \citenamefont {Georgii}, \citenamefont {B{\"o}ni},\ and\
  \citenamefont {Pfleiderer}}]{janoschek2013}%
  \BibitemOpen
  \bibfield  {author} {\bibinfo {author} {\bibfnamefont {M.}~\bibnamefont
  {Janoschek}}, \bibinfo {author} {\bibfnamefont {M.}~\bibnamefont {Garst}},
  \bibinfo {author} {\bibfnamefont {A.}~\bibnamefont {Bauer}}, \bibinfo
  {author} {\bibfnamefont {P.}~\bibnamefont {Krautscheid}}, \bibinfo {author}
  {\bibfnamefont {R.}~\bibnamefont {Georgii}}, \bibinfo {author} {\bibfnamefont
  {P.}~\bibnamefont {B{\"o}ni}}, \ and\ \bibinfo {author} {\bibfnamefont
  {C.}~\bibnamefont {Pfleiderer}},\ }\href@noop {} {\bibfield  {journal}
  {\bibinfo  {journal} {Phys. Rev. B}\ }\textbf {\bibinfo {volume} {87}},\
  \bibinfo {pages} {134407} (\bibinfo {year} {2013})}\BibitemShut {NoStop}%
\bibitem [{\citenamefont {Grigoriev}()}]{Grigoriev_private}%
  \BibitemOpen
  \bibfield  {author} {\bibinfo {author} {\bibfnamefont {S.~V.}\ \bibnamefont
  {Grigoriev}},\ }\href@noop {} {}\bibinfo {howpublished} {private
  communication}\BibitemShut {NoStop}%
\bibitem [{\citenamefont {Grigoriev}\ \emph {et~al.}(2010)\citenamefont
  {Grigoriev}, \citenamefont {Chernyshov}, \citenamefont {Dyadkin},
  \citenamefont {Dmitriev}, \citenamefont {Moskvin}, \citenamefont {Lamago},
  \citenamefont {Wolf}, \citenamefont {Menzel}, \citenamefont {Schoenes},
  \citenamefont {Maleyev},\ and\ \citenamefont
  {Eckerlebe}}]{Grigoriev2010_MnFeSi_chiral}%
  \BibitemOpen
  \bibfield  {author} {\bibinfo {author} {\bibfnamefont {S.~V.}\ \bibnamefont
  {Grigoriev}}, \bibinfo {author} {\bibfnamefont {D.}~\bibnamefont
  {Chernyshov}}, \bibinfo {author} {\bibfnamefont {V.~A.}\ \bibnamefont
  {Dyadkin}}, \bibinfo {author} {\bibfnamefont {V.}~\bibnamefont {Dmitriev}},
  \bibinfo {author} {\bibfnamefont {E.~V.}\ \bibnamefont {Moskvin}}, \bibinfo
  {author} {\bibfnamefont {D.}~\bibnamefont {Lamago}}, \bibinfo {author}
  {\bibfnamefont {T.}~\bibnamefont {Wolf}}, \bibinfo {author} {\bibfnamefont
  {D.}~\bibnamefont {Menzel}}, \bibinfo {author} {\bibfnamefont
  {J.}~\bibnamefont {Schoenes}}, \bibinfo {author} {\bibfnamefont {S.~V.}\
  \bibnamefont {Maleyev}}, \ and\ \bibinfo {author} {\bibfnamefont
  {H.}~\bibnamefont {Eckerlebe}},\ }\href@noop {} {\bibfield  {journal}
  {\bibinfo  {journal} {Phys. Rev. B}\ }\textbf {\bibinfo {volume} {81}},\
  \bibinfo {pages} {012408} (\bibinfo {year} {2010})}\BibitemShut {NoStop}%
\bibitem [{\citenamefont {Bannenberg}\ \emph {et~al.}(2017)\citenamefont
  {Bannenberg}, \citenamefont {Kakurai}, \citenamefont {Falus}, \citenamefont
  {Leli{\`e}vre-Berna}, \citenamefont {Dalgliesh}, \citenamefont {Dewhurst},
  \citenamefont {Qian}, \citenamefont {Onose}, \citenamefont {Endoh},
  \citenamefont {Tokura},\ and\ \citenamefont {Pappas}}]{bannenberg2017}%
  \BibitemOpen
  \bibfield  {author} {\bibinfo {author} {\bibfnamefont {L.~J.}\ \bibnamefont
  {Bannenberg}}, \bibinfo {author} {\bibfnamefont {K.}~\bibnamefont {Kakurai}},
  \bibinfo {author} {\bibfnamefont {P.}~\bibnamefont {Falus}}, \bibinfo
  {author} {\bibfnamefont {E.}~\bibnamefont {Leli{\`e}vre-Berna}}, \bibinfo
  {author} {\bibfnamefont {R.~M.}\ \bibnamefont {Dalgliesh}}, \bibinfo {author}
  {\bibfnamefont {C.~D.}\ \bibnamefont {Dewhurst}}, \bibinfo {author}
  {\bibfnamefont {F.}~\bibnamefont {Qian}}, \bibinfo {author} {\bibfnamefont
  {Y.}~\bibnamefont {Onose}}, \bibinfo {author} {\bibfnamefont
  {Y.}~\bibnamefont {Endoh}}, \bibinfo {author} {\bibfnamefont
  {Y.}~\bibnamefont {Tokura}}, \ and\ \bibinfo {author} {\bibfnamefont
  {C.}~\bibnamefont {Pappas}},\ }\href@noop {} {\bibfield  {journal} {\bibinfo
  {journal} {Phys. Rev. B}\ }\textbf {\bibinfo {volume} {95}},\ \bibinfo
  {pages} {144433} (\bibinfo {year} {2017})}\BibitemShut {NoStop}%
\bibitem [{\citenamefont {Bannenberg}\ \emph {et~al.}(2016)\citenamefont
  {Bannenberg}, \citenamefont {Lefering}, \citenamefont {Kakurai},
  \citenamefont {Onose}, \citenamefont {Endoh}, \citenamefont {Tokura},\ and\
  \citenamefont {Pappas}}]{bannenberg2016squid}%
  \BibitemOpen
  \bibfield  {author} {\bibinfo {author} {\bibfnamefont {L.~J.}\ \bibnamefont
  {Bannenberg}}, \bibinfo {author} {\bibfnamefont {A.~J.~E.}\ \bibnamefont
  {Lefering}}, \bibinfo {author} {\bibfnamefont {K.}~\bibnamefont {Kakurai}},
  \bibinfo {author} {\bibfnamefont {Y.}~\bibnamefont {Onose}}, \bibinfo
  {author} {\bibfnamefont {Y.}~\bibnamefont {Endoh}}, \bibinfo {author}
  {\bibfnamefont {Y.}~\bibnamefont {Tokura}}, \ and\ \bibinfo {author}
  {\bibfnamefont {C.}~\bibnamefont {Pappas}},\ }\href@noop {} {\bibfield
  {journal} {\bibinfo  {journal} {Phys. Rev. B}\ }\textbf {\bibinfo {volume}
  {94}},\ \bibinfo {pages} {134433} (\bibinfo {year} {2016})}\BibitemShut
  {NoStop}%
\bibitem [{\citenamefont {Birch}\ \emph {et~al.}(2019)\citenamefont {Birch},
  \citenamefont {Takagi}, \citenamefont {Seki}, \citenamefont {Wilson},
  \citenamefont {Kagawa}, \citenamefont {{\v{S}}tefan{\v c}i{\v c}},
  \citenamefont {Balakrishnan}, \citenamefont {Fan}, \citenamefont {Steadman},
  \citenamefont {Ottley}, \citenamefont {Crisanti}, \citenamefont {Cubitt},
  \citenamefont {Lancaster}, \citenamefont {Tokura},\ and\ \citenamefont
  {Hatton}}]{Birch2019dw}%
  \BibitemOpen
  \bibfield  {author} {\bibinfo {author} {\bibfnamefont {M.~T.}\ \bibnamefont
  {Birch}}, \bibinfo {author} {\bibfnamefont {R.}~\bibnamefont {Takagi}},
  \bibinfo {author} {\bibfnamefont {S.}~\bibnamefont {Seki}}, \bibinfo {author}
  {\bibfnamefont {M.~N.}\ \bibnamefont {Wilson}}, \bibinfo {author}
  {\bibfnamefont {F.}~\bibnamefont {Kagawa}}, \bibinfo {author} {\bibfnamefont
  {A.}~\bibnamefont {{\v{S}}tefan{\v c}i{\v c}}}, \bibinfo {author}
  {\bibfnamefont {G.}~\bibnamefont {Balakrishnan}}, \bibinfo {author}
  {\bibfnamefont {R.}~\bibnamefont {Fan}}, \bibinfo {author} {\bibfnamefont
  {P.}~\bibnamefont {Steadman}}, \bibinfo {author} {\bibfnamefont {C.~J.}\
  \bibnamefont {Ottley}}, \bibinfo {author} {\bibfnamefont {M.}~\bibnamefont
  {Crisanti}}, \bibinfo {author} {\bibfnamefont {R.}~\bibnamefont {Cubitt}},
  \bibinfo {author} {\bibfnamefont {T.}~\bibnamefont {Lancaster}}, \bibinfo
  {author} {\bibfnamefont {Y.}~\bibnamefont {Tokura}}, \ and\ \bibinfo {author}
  {\bibfnamefont {P.~D.}\ \bibnamefont {Hatton}},\ }\href@noop {} {\bibfield
  {journal} {\bibinfo  {journal} {Phys. Rev. B}\ }\textbf {\bibinfo {volume}
  {100}},\ \bibinfo {pages} {014425} (\bibinfo {year} {2019})}\BibitemShut
  {NoStop}%
\bibitem [{\citenamefont {Pickup}\ \emph {et~al.}(2009)\citenamefont {Pickup},
  \citenamefont {Cywinski}, \citenamefont {Pappas}, \citenamefont {Farago},\
  and\ \citenamefont {Fouquet}}]{Pickup2009}%
  \BibitemOpen
  \bibfield  {author} {\bibinfo {author} {\bibfnamefont {R.~M.}\ \bibnamefont
  {Pickup}}, \bibinfo {author} {\bibfnamefont {R.}~\bibnamefont {Cywinski}},
  \bibinfo {author} {\bibfnamefont {C.}~\bibnamefont {Pappas}}, \bibinfo
  {author} {\bibfnamefont {B.}~\bibnamefont {Farago}}, \ and\ \bibinfo {author}
  {\bibfnamefont {P.}~\bibnamefont {Fouquet}},\ }\href@noop {} {\bibfield
  {journal} {\bibinfo  {journal} {Phys. Rev. Letters}\ }\textbf {\bibinfo
  {volume} {102}},\ \bibinfo {pages} {097202} (\bibinfo {year}
  {2009})}\BibitemShut {NoStop}%
\bibitem [{\citenamefont {Pappas}\ \emph {et~al.}(2003)\citenamefont {Pappas},
  \citenamefont {Mezei}, \citenamefont {Ehlers}, \citenamefont {Manuel},\ and\
  \citenamefont {Campbell}}]{Pappas2003}%
  \BibitemOpen
  \bibfield  {author} {\bibinfo {author} {\bibfnamefont {C.}~\bibnamefont
  {Pappas}}, \bibinfo {author} {\bibfnamefont {F.}~\bibnamefont {Mezei}},
  \bibinfo {author} {\bibfnamefont {G.}~\bibnamefont {Ehlers}}, \bibinfo
  {author} {\bibfnamefont {P.}~\bibnamefont {Manuel}}, \ and\ \bibinfo {author}
  {\bibfnamefont {I.~A.}\ \bibnamefont {Campbell}},\ }\href@noop {} {\bibfield
  {journal} {\bibinfo  {journal} {Phys. Rev. B}\ }\textbf {\bibinfo {volume}
  {68}},\ \bibinfo {pages} {054431} (\bibinfo {year} {2003})}\BibitemShut
  {NoStop}%
\bibitem [{\citenamefont {Dzyaloshinskii}(1964)}]{Dzyaloshinskii1964a}%
  \BibitemOpen
  \bibfield  {author} {\bibinfo {author} {\bibfnamefont {I.~E.}\ \bibnamefont
  {Dzyaloshinskii}},\ }\href@noop {} {\bibfield  {journal} {\bibinfo  {journal}
  {Soviet Physics JETP}\ }\textbf {\bibinfo {volume} {19}},\ \bibinfo {pages}
  {960} (\bibinfo {year} {1964})}\BibitemShut {NoStop}%
\bibitem [{\citenamefont {Dzyaloshinskii}(1965)}]{Dzyaloshinskii1965a}%
  \BibitemOpen
  \bibfield  {author} {\bibinfo {author} {\bibfnamefont {I.~E.}\ \bibnamefont
  {Dzyaloshinskii}},\ }\href@noop {} {\bibfield  {journal} {\bibinfo  {journal}
  {Soviet Physics JETP}\ }\textbf {\bibinfo {volume} {20}},\ \bibinfo {pages}
  {665} (\bibinfo {year} {1965})}\BibitemShut {NoStop}%
\bibitem [{\citenamefont {Bak}\ and\ \citenamefont {Jensen}(1980)}]{bak1980}%
  \BibitemOpen
  \bibfield  {author} {\bibinfo {author} {\bibfnamefont {P.}~\bibnamefont
  {Bak}}\ and\ \bibinfo {author} {\bibfnamefont {M.~H.}\ \bibnamefont
  {Jensen}},\ }\href@noop {} {\bibfield  {journal} {\bibinfo  {journal}
  {Journal of Physics C: Solid State Physics}\ }\textbf {\bibinfo {volume}
  {13}},\ \bibinfo {pages} {L881} (\bibinfo {year} {1980})}\BibitemShut
  {NoStop}%
\bibitem [{\citenamefont {Qian}\ \emph {et~al.}(2018)\citenamefont {Qian},
  \citenamefont {Bannenberg}, \citenamefont {Wilhelm}, \citenamefont
  {Chaboussant}, \citenamefont {DeBeer-Schmitt}, \citenamefont {Schmidt},
  \citenamefont {Aqeel}, \citenamefont {Palstra}, \citenamefont {Br{\"u}ck},
  \citenamefont {Lefering}, \citenamefont {Pappas}, \citenamefont {Mostovoy},\
  and\ \citenamefont {Leonov}}]{qian2018}%
  \BibitemOpen
  \bibfield  {author} {\bibinfo {author} {\bibfnamefont {F.}~\bibnamefont
  {Qian}}, \bibinfo {author} {\bibfnamefont {L.~J.}\ \bibnamefont
  {Bannenberg}}, \bibinfo {author} {\bibfnamefont {H.}~\bibnamefont {Wilhelm}},
  \bibinfo {author} {\bibfnamefont {G.}~\bibnamefont {Chaboussant}}, \bibinfo
  {author} {\bibfnamefont {L.~M.}\ \bibnamefont {DeBeer-Schmitt}}, \bibinfo
  {author} {\bibfnamefont {M.~P.}\ \bibnamefont {Schmidt}}, \bibinfo {author}
  {\bibfnamefont {A.}~\bibnamefont {Aqeel}}, \bibinfo {author} {\bibfnamefont
  {T.~T.~M.}\ \bibnamefont {Palstra}}, \bibinfo {author} {\bibfnamefont
  {E.~H.}\ \bibnamefont {Br{\"u}ck}}, \bibinfo {author} {\bibfnamefont
  {A.~J.~E.}\ \bibnamefont {Lefering}}, \bibinfo {author} {\bibfnamefont
  {C.}~\bibnamefont {Pappas}}, \bibinfo {author} {\bibfnamefont
  {M.}~\bibnamefont {Mostovoy}}, \ and\ \bibinfo {author} {\bibfnamefont
  {A.~O.}\ \bibnamefont {Leonov}},\ }\href@noop {} {\bibfield  {journal}
  {\bibinfo  {journal} {Science Advances}\ }\textbf {\bibinfo {volume} {4}},\
  \bibinfo {pages} {eaat7323} (\bibinfo {year} {2018})}\BibitemShut {NoStop}%
\bibitem [{\citenamefont {Tanigaki}\ \emph {et~al.}(2015)\citenamefont
  {Tanigaki}, \citenamefont {Shibata}, \citenamefont {Kanazawa}, \citenamefont
  {Yu}, \citenamefont {Y.~Onose}, \citenamefont {Park}, \citenamefont
  {Shindo},\ and\ \citenamefont {Tokura}}]{TanigakiNanoLett}%
  \BibitemOpen
  \bibfield  {author} {\bibinfo {author} {\bibfnamefont {T.}~\bibnamefont
  {Tanigaki}}, \bibinfo {author} {\bibfnamefont {K.}~\bibnamefont {Shibata}},
  \bibinfo {author} {\bibfnamefont {N.}~\bibnamefont {Kanazawa}}, \bibinfo
  {author} {\bibfnamefont {X.}~\bibnamefont {Yu}}, \bibinfo {author}
  {\bibfnamefont {Y.}~\bibnamefont {Y.~Onose}}, \bibinfo {author}
  {\bibfnamefont {H.~S.}\ \bibnamefont {Park}}, \bibinfo {author}
  {\bibfnamefont {D.}~\bibnamefont {Shindo}}, \ and\ \bibinfo {author}
  {\bibfnamefont {Y.}~\bibnamefont {Tokura}},\ }\href@noop {} {\bibfield
  {journal} {\bibinfo  {journal} {Nano Letters}\ }\textbf {\bibinfo {volume}
  {15}},\ \bibinfo {pages} {5438} (\bibinfo {year} {2015})}\BibitemShut
  {NoStop}%
\bibitem [{\citenamefont {Fujishiro}\ \emph {et~al.}(2019)\citenamefont
  {Fujishiro}, \citenamefont {Kanazawa}, \citenamefont {Nakajima},
  \citenamefont {Yu}, \citenamefont {Ohishi}, \citenamefont {Kawamura},
  \citenamefont {Kakurai}, \citenamefont {Arima}, \citenamefont {Mitamura},
  \citenamefont {Miyake}, \citenamefont {Akiba}, \citenamefont {Tokunaga},
  \citenamefont {Matsuo}, \citenamefont {Kindo}, \citenamefont {Koretsune},
  \citenamefont {Arita},\ and\ \citenamefont {Tokura}}]{FujishiroNatCommun}%
  \BibitemOpen
  \bibfield  {author} {\bibinfo {author} {\bibfnamefont {Y.}~\bibnamefont
  {Fujishiro}}, \bibinfo {author} {\bibfnamefont {N.}~\bibnamefont {Kanazawa}},
  \bibinfo {author} {\bibfnamefont {T.}~\bibnamefont {Nakajima}}, \bibinfo
  {author} {\bibfnamefont {X.~Z.}\ \bibnamefont {Yu}}, \bibinfo {author}
  {\bibfnamefont {K.}~\bibnamefont {Ohishi}}, \bibinfo {author} {\bibfnamefont
  {Y.}~\bibnamefont {Kawamura}}, \bibinfo {author} {\bibfnamefont
  {K.}~\bibnamefont {Kakurai}}, \bibinfo {author} {\bibfnamefont
  {T.}~\bibnamefont {Arima}}, \bibinfo {author} {\bibfnamefont
  {H.}~\bibnamefont {Mitamura}}, \bibinfo {author} {\bibfnamefont
  {A.}~\bibnamefont {Miyake}}, \bibinfo {author} {\bibfnamefont
  {K.}~\bibnamefont {Akiba}}, \bibinfo {author} {\bibfnamefont
  {M.}~\bibnamefont {Tokunaga}}, \bibinfo {author} {\bibfnamefont
  {A.}~\bibnamefont {Matsuo}}, \bibinfo {author} {\bibfnamefont
  {K.}~\bibnamefont {Kindo}}, \bibinfo {author} {\bibfnamefont
  {T.}~\bibnamefont {Koretsune}}, \bibinfo {author} {\bibfnamefont
  {R.}~\bibnamefont {Arita}}, \ and\ \bibinfo {author} {\bibfnamefont
  {Y.}~\bibnamefont {Tokura}},\ }\href@noop {} {\bibfield  {journal} {\bibinfo
  {journal} {Nature Communications}\ }\textbf {\bibinfo {volume} {10}},\
  \bibinfo {pages} {1059} (\bibinfo {year} {2019})}\BibitemShut {NoStop}%
\bibitem [{\citenamefont {Mutter}\ \emph {et~al.}(2019)\citenamefont {Mutter},
  \citenamefont {Leonov},\ and\ \citenamefont {Inoue}}]{Mutter2019}%
  \BibitemOpen
  \bibfield  {author} {\bibinfo {author} {\bibfnamefont {T.~T.~J.}\
  \bibnamefont {Mutter}}, \bibinfo {author} {\bibfnamefont {A.~O.}\
  \bibnamefont {Leonov}}, \ and\ \bibinfo {author} {\bibfnamefont
  {K.}~\bibnamefont {Inoue}},\ }\href@noop {} {\bibfield  {journal} {\bibinfo
  {journal} {Phys. Rev. B}\ }\textbf {\bibinfo {volume} {100}},\ \bibinfo
  {pages} {060407(R)} (\bibinfo {year} {2019})}\BibitemShut {NoStop}%
\bibitem [{\citenamefont {Skyrme}(1961)}]{skyrme1961}%
  \BibitemOpen
  \bibfield  {author} {\bibinfo {author} {\bibfnamefont {T.~H.~R.}\
  \bibnamefont {Skyrme}},\ }\href@noop {} {\bibfield  {journal} {\bibinfo
  {journal} {Proceedings of the Royal Society Londen A}\ }\textbf {\bibinfo
  {volume} {260}},\ \bibinfo {pages} {127} (\bibinfo {year}
  {1961})}\BibitemShut {NoStop}%
\bibitem [{\citenamefont {Leonov}\ and\ \citenamefont
  {Mostovoy}(2015)}]{Leonov_Mostovoy_NatCom_2015}%
  \BibitemOpen
  \bibfield  {author} {\bibinfo {author} {\bibfnamefont {A.~O.}\ \bibnamefont
  {Leonov}}\ and\ \bibinfo {author} {\bibfnamefont {M.}~\bibnamefont
  {Mostovoy}},\ }\href@noop {} {\bibfield  {journal} {\bibinfo  {journal}
  {Nature Communications}\ }\textbf {\bibinfo {volume} {6}},\ \bibinfo {pages}
  {8275} (\bibinfo {year} {2015})}\BibitemShut {NoStop}%
\bibitem [{\citenamefont {Demishev}\ \emph {et~al.}(2013)\citenamefont
  {Demishev}, \citenamefont {Lobanova}, \citenamefont {Glushkov}, \citenamefont
  {Ischenko}, \citenamefont {Sluchanko}, \citenamefont {Dyadkin}, \citenamefont
  {Potapova},\ and\ \citenamefont {Grigoriev}}]{demishev2013}%
  \BibitemOpen
  \bibfield  {author} {\bibinfo {author} {\bibfnamefont {S.~V.}\ \bibnamefont
  {Demishev}}, \bibinfo {author} {\bibfnamefont {I.~I.}\ \bibnamefont
  {Lobanova}}, \bibinfo {author} {\bibfnamefont {V.~V.}\ \bibnamefont
  {Glushkov}}, \bibinfo {author} {\bibfnamefont {T.~V.}\ \bibnamefont
  {Ischenko}}, \bibinfo {author} {\bibfnamefont {N.~E.}\ \bibnamefont
  {Sluchanko}}, \bibinfo {author} {\bibfnamefont {V.~A.}\ \bibnamefont
  {Dyadkin}}, \bibinfo {author} {\bibfnamefont {N.~M.}\ \bibnamefont
  {Potapova}}, \ and\ \bibinfo {author} {\bibfnamefont {S.~V.}\ \bibnamefont
  {Grigoriev}},\ }\href@noop {} {\bibfield  {journal} {\bibinfo  {journal}
  {JETP Letters}\ }\textbf {\bibinfo {volume} {98}},\ \bibinfo {pages} {829}
  (\bibinfo {year} {2013})}\BibitemShut {NoStop}%
\bibitem [{\citenamefont {Glushkov}\ \emph {et~al.}(2015)\citenamefont
  {Glushkov}, \citenamefont {Lobanova}, \citenamefont {Ivanov}, \citenamefont
  {Voronov}, \citenamefont {Dyadkin}, \citenamefont {Chubova}, \citenamefont
  {Grigoriev},\ and\ \citenamefont {Demishev}}]{glushkov2015}%
  \BibitemOpen
  \bibfield  {author} {\bibinfo {author} {\bibfnamefont {V.~V.}\ \bibnamefont
  {Glushkov}}, \bibinfo {author} {\bibfnamefont {I.~I.}\ \bibnamefont
  {Lobanova}}, \bibinfo {author} {\bibfnamefont {V.~Y.}\ \bibnamefont
  {Ivanov}}, \bibinfo {author} {\bibfnamefont {V.~V.}\ \bibnamefont {Voronov}},
  \bibinfo {author} {\bibfnamefont {V.~A.}\ \bibnamefont {Dyadkin}}, \bibinfo
  {author} {\bibfnamefont {N.~M.}\ \bibnamefont {Chubova}}, \bibinfo {author}
  {\bibfnamefont {S.~V.}\ \bibnamefont {Grigoriev}}, \ and\ \bibinfo {author}
  {\bibfnamefont {S.~V.}\ \bibnamefont {Demishev}},\ }\href@noop {} {\bibfield
  {journal} {\bibinfo  {journal} {Phys. Rev. Lett.}\ }\textbf {\bibinfo
  {volume} {115}},\ \bibinfo {pages} {256601} (\bibinfo {year}
  {2015})}\BibitemShut {NoStop}%
\bibitem [{\citenamefont {Kurumaji}\ \emph {et~al.}(2019)\citenamefont
  {Kurumaji}, \citenamefont {Nakajima}, \citenamefont {Hirschberger},
  \citenamefont {Kikkawa}, \citenamefont {Yamasaki}, \citenamefont {Sagayama},
  \citenamefont {Nakao}, \citenamefont {Taguchi}, \citenamefont {Arima},\ and\
  \citenamefont {Tokura}}]{2019_Science_frustr_SKL}%
  \BibitemOpen
  \bibfield  {author} {\bibinfo {author} {\bibfnamefont {T.}~\bibnamefont
  {Kurumaji}}, \bibinfo {author} {\bibfnamefont {T.}~\bibnamefont {Nakajima}},
  \bibinfo {author} {\bibfnamefont {M.}~\bibnamefont {Hirschberger}}, \bibinfo
  {author} {\bibfnamefont {A.}~\bibnamefont {Kikkawa}}, \bibinfo {author}
  {\bibfnamefont {Y.}~\bibnamefont {Yamasaki}}, \bibinfo {author}
  {\bibfnamefont {H.}~\bibnamefont {Sagayama}}, \bibinfo {author}
  {\bibfnamefont {H.}~\bibnamefont {Nakao}}, \bibinfo {author} {\bibfnamefont
  {Y.}~\bibnamefont {Taguchi}}, \bibinfo {author} {\bibfnamefont {T.-h.}\
  \bibnamefont {Arima}}, \ and\ \bibinfo {author} {\bibfnamefont
  {Y.}~\bibnamefont {Tokura}},\ }\href@noop {} {\bibfield  {journal} {\bibinfo
  {journal} {Science}\ }\textbf {\bibinfo {volume} {365}},\ \bibinfo {pages}
  {914} (\bibinfo {year} {2019})}\BibitemShut {NoStop}%
\bibitem [{\citenamefont {von Malottki}\ \emph {et~al.}(2019)\citenamefont {von
  Malottki}, \citenamefont {Dup{\'e}}, \citenamefont {Bessarab}, \citenamefont
  {Delin},\ and\ \citenamefont {Heinze}}]{vonMalottki_Heinze_2019}%
  \BibitemOpen
  \bibfield  {author} {\bibinfo {author} {\bibfnamefont {S.}~\bibnamefont {von
  Malottki}}, \bibinfo {author} {\bibfnamefont {B.}~\bibnamefont {Dup{\'e}}},
  \bibinfo {author} {\bibfnamefont {P.~F.}\ \bibnamefont {Bessarab}}, \bibinfo
  {author} {\bibfnamefont {A.}~\bibnamefont {Delin}}, \ and\ \bibinfo {author}
  {\bibfnamefont {S.}~\bibnamefont {Heinze}},\ }\href@noop {} {\bibfield
  {journal} {\bibinfo  {journal} {Scientific reports}\ }\textbf {\bibinfo
  {volume} {9}},\ \bibinfo {pages} {12299} (\bibinfo {year}
  {2019})}\BibitemShut {NoStop}%
\bibitem [{\citenamefont {Yuan}\ \emph {et~al.}(2017)\citenamefont {Yuan},
  \citenamefont {Gomonay},\ and\ \citenamefont {Kl{\"a}ui}}]{Yuan_2017_Klaui}%
  \BibitemOpen
  \bibfield  {author} {\bibinfo {author} {\bibfnamefont {H.~Y.}\ \bibnamefont
  {Yuan}}, \bibinfo {author} {\bibfnamefont {O.}~\bibnamefont {Gomonay}}, \
  and\ \bibinfo {author} {\bibfnamefont {M.}~\bibnamefont {Kl{\"a}ui}},\
  }\href@noop {} {\bibfield  {journal} {\bibinfo  {journal} {Phys. Rev B}\
  }\textbf {\bibinfo {volume} {96}},\ \bibinfo {pages} {134415} (\bibinfo
  {year} {2017})}\BibitemShut {NoStop}%
\bibitem [{\citenamefont {Laliena}\ \emph {et~al.}(2016)\citenamefont
  {Laliena}, \citenamefont {Campo},\ and\ \citenamefont
  {Kousaka}}]{Laliena_Campo_2016}%
  \BibitemOpen
  \bibfield  {author} {\bibinfo {author} {\bibfnamefont {V.}~\bibnamefont
  {Laliena}}, \bibinfo {author} {\bibfnamefont {J.}~\bibnamefont {Campo}}, \
  and\ \bibinfo {author} {\bibfnamefont {Y.}~\bibnamefont {Kousaka}},\
  }\href@noop {} {\bibfield  {journal} {\bibinfo  {journal} {Phys. Rev B}\
  }\textbf {\bibinfo {volume} {94}},\ \bibinfo {pages} {094439} (\bibinfo
  {year} {2016})}\BibitemShut {NoStop}%
\bibitem [{\citenamefont {Leonov}\ and\ \citenamefont
  {Bogdanov}(2018)}]{leonov2018b}%
  \BibitemOpen
  \bibfield  {author} {\bibinfo {author} {\bibfnamefont {A.~O.}\ \bibnamefont
  {Leonov}}\ and\ \bibinfo {author} {\bibfnamefont {A.~N.}\ \bibnamefont
  {Bogdanov}},\ }\href@noop {} {\bibfield  {journal} {\bibinfo  {journal} {New
  Journal of Physics}\ }\textbf {\bibinfo {volume} {20}},\ \bibinfo {pages}
  {043017} (\bibinfo {year} {2018})}\BibitemShut {NoStop}%
\bibitem [{\citenamefont {Shinozaki}\ \emph {et~al.}(2017)\citenamefont
  {Shinozaki}, \citenamefont {Hoshino}, \citenamefont {Masaki}, \citenamefont
  {Bogdanov}, \citenamefont {Leonov}, \citenamefont {Kishine},\ and\
  \citenamefont {Kato}}]{Shinozaki_Leonov_2017}%
  \BibitemOpen
  \bibfield  {author} {\bibinfo {author} {\bibfnamefont {M.}~\bibnamefont
  {Shinozaki}}, \bibinfo {author} {\bibfnamefont {S.}~\bibnamefont {Hoshino}},
  \bibinfo {author} {\bibfnamefont {Y.}~\bibnamefont {Masaki}}, \bibinfo
  {author} {\bibfnamefont {A.~N.}\ \bibnamefont {Bogdanov}}, \bibinfo {author}
  {\bibfnamefont {A.~O.}\ \bibnamefont {Leonov}}, \bibinfo {author}
  {\bibfnamefont {J.-I.}\ \bibnamefont {Kishine}}, \ and\ \bibinfo {author}
  {\bibfnamefont {Y.}~\bibnamefont {Kato}},\ }\href@noop {} {\bibfield
  {journal} {\bibinfo  {journal} {arXiv.org}\ ,\ \bibinfo {pages}
  {arXiv:1705.07778}} (\bibinfo {year} {2017})},\ \Eprint
  {http://arxiv.org/abs/1705.07778} {1705.07778} \BibitemShut {NoStop}%
\bibitem [{\citenamefont {Kharkov}\ \emph {et~al.}(2017)\citenamefont
  {Kharkov}, \citenamefont {Sushkov},\ and\ \citenamefont
  {Mostovoy}}]{Kharkov_Maxim_2017}%
  \BibitemOpen
  \bibfield  {author} {\bibinfo {author} {\bibfnamefont {Y.~A.}\ \bibnamefont
  {Kharkov}}, \bibinfo {author} {\bibfnamefont {O.~P.}\ \bibnamefont
  {Sushkov}}, \ and\ \bibinfo {author} {\bibfnamefont {M.}~\bibnamefont
  {Mostovoy}},\ }\href@noop {} {\bibfield  {journal} {\bibinfo  {journal}
  {Phys. Rev. Lett.}\ }\textbf {\bibinfo {volume} {119}},\ \bibinfo {pages}
  {207201} (\bibinfo {year} {2017})}\BibitemShut {NoStop}%
\bibitem [{\citenamefont {Schaub}\ and\ \citenamefont
  {Mukamel}(1985)}]{Mukamel_1985}%
  \BibitemOpen
  \bibfield  {author} {\bibinfo {author} {\bibfnamefont {B.}~\bibnamefont
  {Schaub}}\ and\ \bibinfo {author} {\bibfnamefont {D.}~\bibnamefont
  {Mukamel}},\ }\href@noop {} {\bibfield  {journal} {\bibinfo  {journal} {Phys.
  Rev B}\ }\textbf {\bibinfo {volume} {32}},\ \bibinfo {pages} {6385} (\bibinfo
  {year} {1985})}\BibitemShut {NoStop}%
\end{thebibliography}

%

\end{document}